\documentclass[11pt,letterpaper]{article}

\usepackage[T1]{fontenc}
\usepackage[english]{babel}
\usepackage{times}

\usepackage[letterpaper,margin=1in]{geometry}
\usepackage[parfill]{parskip}
\usepackage{setspace}

\usepackage{amsmath,amstext,amsbsy,amsopn,amsthm,amsfonts,amssymb}
\usepackage{booktabs,array,colortbl}
\usepackage[]{graphicx}
\usepackage{rotate}
\usepackage{wrapfig}
\usepackage{caption}
\usepackage{subfigure}
\usepackage{lastpage}

\usepackage[small,compact]{titlesec}
\usepackage{paralist}

\usepackage{verbatim} 
\usepackage[usenames,dvipsnames,svgnames]{xcolor}
\usepackage{tikz} 
\usepackage{pgfgantt}
\usepackage{pgfcalendar}
\usetikzlibrary{shadows}
\usetikzlibrary{shadings}

\usepackage{fancyhdr}
\renewenvironment{thebibliography}[1]{%
    \begin{oldthebibliography}{#1}%
    \setlength{\parskip}{0.0cm}%
    \setlength{\itemsep}{0.0cm}%
    }%
    {%
        \end{oldthebibliography}%
    }

\begin{document}

\title{\textbf{ISWAR}: An \textbf{I}maging \textbf{S}ystem with \textbf{W}atermarking \\ and \textbf{A}ttack \textbf{R}esilience}

\maketitle

\begin{center}
\begin{tabular}{c}
Saraju P. Mohanty \\
NanoSystem Design Laboratory (NSDL)
(http://nsdl.cse.unt.edu)\\
Department of Computer Science and Engineering \\
University of North Texas, Denton, TX 76203.\\
Email : saraju.mohanty@unt.edu 
\end{tabular}
\end{center}

\cfoot{Page -- \thepage-of-\pageref{LastPage}}

\begin{abstract}
With the explosive growth of internet technology, easy transfer of
digital multimedia is feasible.
However, this kind of convenience with which authorized
users can access information, turns out to be a mixed blessing
due to information piracy. The emerging field of Digital
Rights Management (DRM) systems addresses issues related to
the intellectual property rights of digital content. In this paper,
an object-oriented (OO) DRM system, called
``\textbf{I}maging \textbf{S}ystem with \textbf{W}atermarking and
\textbf{A}ttack \textbf{R}esilience'' (ISWAR), is presented that generates and
authenticates color images with embedded mechanisms for protection
against infringement of ownership rights as well as security
attacks. In addition to the methods, in the object-oriented sense, for performing
traditional encryption and decryption, the system implements methods
for visible and invisible watermarking. This paper presents one
visible and one invisible watermarking algorithm that have been
integrated in the system. The qualitative and
quantitative results obtained for these two watermarking algorithms
with several benchmark images
indicate that high-quality watermarked images are produced by
the algorithms. With the help of experimental
results it is demonstrated that the
presented invisible watermarking techniques are resilient to
the well known benchmark attacks and hence a fail-safe
method for providing constant protection to ownership rights.
\end{abstract}

\newline

\section{Introduction}
\label{Introduction}

The Internet revolution towards the end of the last millennium, has ushered
in a new era of information technology. There has been an explosive
growth in multimedia applications including video-on-demand,
and distant education. An interesting trend of this new era
is to move from conventional to digital libraries. This, in turn,
provides  tremendous opportunities for scholarly research and
education because of the ease with which images and texts can be
availed through the internet. However, this kind of ultimate
flexibility to avail digital content, particularly that of
images,  has also its negative side. Easy access facilitates
information  piracy  through unauthorized replication and
manipulation of digital content with the help of inexpensive tools.
Hence, of late, concerns about protection and enforcement of
intellectual property (IP) rights of the digital content involved in
these transactions, have also been mounting. The emerging field of
digital rights management (DRM) systems \cite{EmmanuelMMSJ2003,
KundurIEEEProceedings2004, MohantyJSA2009Oct} addresses these issues related to
ownership rights of digital content. Specifically, issues addressed
by DRM systems encompass various aspects of content management,
namely content identification, storage, representation,
distribution and intellectual property rights management. Two basic
goals of DRM systems that can be met with digital watermarking are:
(1) preventing unauthorized use of the images (in general, any digital
information), particularly for commercial purposes and  (2) providing
visibility to the authentic source or the owner of the information
on a continuous basis.

Digital watermarking, in essence, is the process of embedding into a
multimedia object a digital signature or data that is variously
known as watermark, tag or label. Detection or extraction of this
watermark at a later juncture enables assertions about
the authenticity and ownership of the object
\cite{KougianosIJCEE2009Mar}. Hence,
watermarking is one of the key technologies that can be used for
establishing ownership rights, tracking usage, ensuring authorized
access, preventing illegal replication, and facilitating content
authentication \cite{BenderIBMSystemsJournal2000, MasonIBC2000, NikolaidisICIP2001, CoxJASP2002, EmmanuelMMSJ2003, KougianosIJCEE2009Mar, MohantyJSA2009Oct}.
In general, a watermarking scheme consists of three
parts, namely: (1) creation or procurement (say, from a library) of
the watermark, (2) insertion or encoding of the watermark, and (3)
detection or extraction and verification of the watermark. The
insertion algorithm incorporates the watermark into the object,
whereas the verification algorithm authenticates the object,
determining both the owner and the integrity of the object. Our
\textbf{I}maging \textbf{S}ystem with \textbf{W}atermarking and
\textbf{A}ttack \textbf{R}esilience (ISWAR) has class methods to perform
all these tasks.

When encryption technologies are used in conjunction with watermarking
\cite{EskiciogluSP2001, MohantyJSA2009Oct}, full protection from unauthorized access of
digital content can be achieved. Our ISWAR precisely uses this
two-tier protection mechanism with methods for traditional
encryption and decryption as well as others for visible and invisible
watermarking. The watermarking methods use the novel algorithms
developed by us for this purpose. Unlike many other secure systems,
our system prevents even authorized users from illegally replicating
the decrypted content. Since it is difficult to alter an image
without damaging the watermark, image integrity can be protected by
watermarking---visible or invisible.

The {\bf novel contributions of this paper} are as follows:
\begin{enumerate}
\item
An object-oriented DRM system, called
``\textbf{I}maging \textbf{S}ystem with \textbf{W}atermarking and
\textbf{A}ttack \textbf{R}esilience'' (ISWAR) is presented. The system generates watermarked color images and authenticates color images for protection
against infringement of ownership rights and security attacks.

\item
A visible watermarking algorithm is presented that adaptively inserts
a watermark on the host color image such that high-quality watermarked color
images are produced from ISWAR.

\item
An invisible watermarking algorithm that uses both encryption and watermarking
for dual-layer protection of the watermark.

\item
Complete implementation of ISAWR imaging system as a user-friendly software.

\item
Exhaustive validation the ISWAR imaging system.

\item
Exhaustive experiments of the visible and invisible watermarking algorithms
with benchmark color images.
\end{enumerate}

The organization of the rest of the paper is as follows.
Section~\ref{sec:TypesWMSurveyPaper} presents the state of the art
in digital watermarking. Against this backdrop, a
high-level view of ISWAR is presented in
Section~\ref{sec:HighLevelViewOfISWAR}.
Sections~\ref{VisibleWatermarkingAlgorithm} and
\ref{InvisibleWatermarkingAlgorithm} present our visible and
invisible watermarking algorithms, respectively. System
implementation, usage and validation are detailed in
section~\ref{sec:ImplementationUsageValidation}.
Section~\ref{sec:AlgoPerformance} presents performance analysis of
the watermarking algorithms. Summary and conclusions are presented in
Section~\ref{sec:SummaryAndConclusion}.

\section{State of the Art of Digital Watermarking}
\label{sec:TypesWMSurveyPaper}

\begin{figure*}[htbp]
\centering
\includegraphics[width=0.96\textwidth] {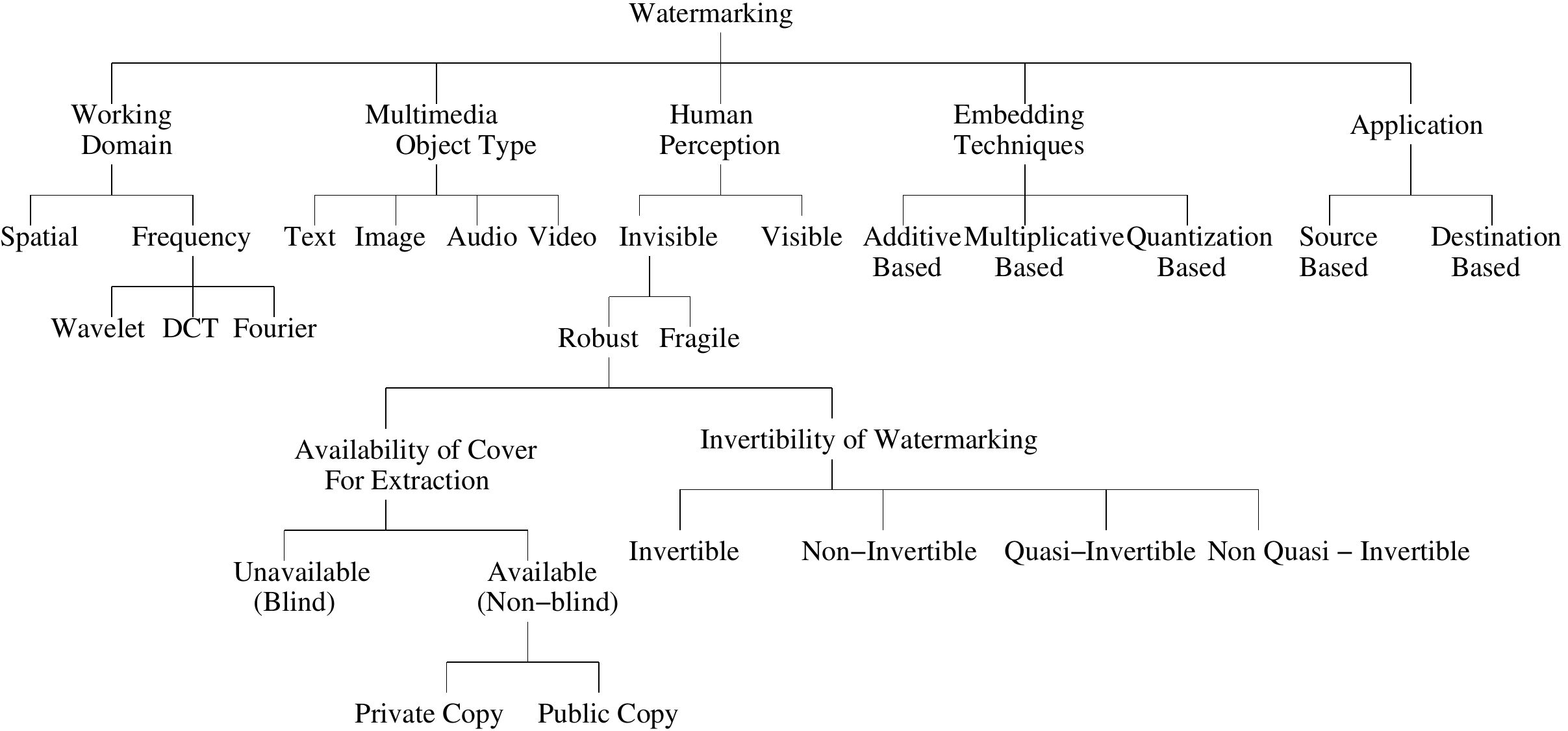}
\caption{Types of watermarking techniques.}
\label{fig:TypesOfDifferentWatermarks}
\end{figure*}

Research on watermarking has matured over the last decade and hence
the current literature abounds with techniques in this area. Various
published works discuss the desired
characteristics of various types of watermarks from the research
point of view as well as from the users' perspective
\cite{BenderIBMSystemsJournal1996, MasonIBC2000, XieSNPD2007, KougianosIJCEE2009Mar}.
The categorization of the watermarking schemes under various
criteria is presented in Figure~\ref{fig:TypesOfDifferentWatermarks}.
At the same time hardware-assisted watermarking approaches are presented
\cite{Mohanty_JSA2009Oct, Mohanty_CDT2007Sep}.

Based on the domain of the cover object used for embedding tokens,
watermarking methods may be classified as spatial domain or
frequency (or transform) domain methods.
Watermarking techniques can be divided into four categories
according to the type of multimedia object to be used, namely,
image, video, audio, or  text watermarking.
Based on human perception, digital watermarks can be divided
into visible and invisible types. In
visible watermarking, which is analogous to watermarking on paper
currency, a secondary translucent image is overlaid on the primary
(cover) image \cite{Mohanty_ICMPS2000}.
The invisible-robust watermark is one that is embedded in such a way
that alterations made to the pixel value are not perceptually noticed
and it can be recovered only with the appropriate decoding
mechanism. Invisible robust watermarks may be further classified on two
criteria --- (1) availability of the original cover image for
extraction/detection purposes and (2) invertibility of the
watermarking scheme. In addition to the above, several hybrid watermarking algorithms
have been proposed in the literature for different purposes
\cite{MohantyACMMM1999, HuaICPRIP2001, SheppardACMMM2001, GuoMMSJ2003}.
Based on the embedding techniques used, watermarking schemes can
be categorized as additive, multiplicative, or quantization
\cite{VoloshynovskiyIEEECommMag2001}.
From the application point of view, digital watermarks could be
classified as source or destination based
\cite{VoyatzisgProceedingsIEEE1999} watermarks.
In summary, there are various currently available
watermarking techniques tailored for the needs of different applications.

Selected literature in watermarking research, whose scope is more close to the research presented in this paper, is discussed
now.
In \cite{SungSERA2006, FrattolilloISIAS2008, LiuCSSE2008},
several software-based DRM and watermarking systems are proposed
which are more or less ready-to-use.
Visible watermarking algorithms which are absolutely essential for
many applications, including protection of publicly available images and video broadcasting, are presented in \cite{BenderIBMSystemsJournal1996, MintzerIBMJournal1996, KankanhalliICMCS1999, MohantyICME2000, LuminiICME2004, HuKwongISCS2004, TopkaraEtAlLecNotesCompSc2005, HuangIEEEMM2006Feb, TsaiICME2007,
ChenCISE2009, YangTCSVT2009May,FarrugiaMELCON2010}.
Invisible-robust watermarking algorithms presented in \cite{ChenICCE2000, WuCGI2001, PaiIEICETIS2006Apr, SaxenaICSPCA2007, RaoIJCSNS2009Mar} hide a binary image inside the host image and hence are well positioned to provide the highest level of image security.

\section{ISWAR -- A High Level View}
\label{sec:HighLevelViewOfISWAR}

This section introduces  a full-fledged
software system called ISWAR that we developed for secure imaging
applications. It seeks to provide the user with controlled access and
data security via data encryption, good network bandwidth
utilization via frequency domain techniques for data reduction, and
copyright protection via watermarking. It can perform both visible
and invisible watermarking of color as well as monochrome images.
This system provides information privacy and resilience to security
attacks by a two layer protection mechanism that is built upon
traditional encryption schemes as well as the novel invisible and
visible watermarking schemes proposed here. Explicitly, the system
was designed to fulfill the following requirements:
\begin{enumerate}
\item
Simplicity
with flexibility and extensibility---With our system, it is
possible to generate images with varied levels of resilience to
attack with visible (e.g., registered company seals) or
robust invisible watermarking with encryption.

\item
Ease in
extraction of watermarks from the images for the purpose of proving
ownership.

\item
No degradation in the cover images due to
watermarking.

\item
High-level of information hiding, i.e.,
the presence of an invisible watermark should not be inferrable from the
statistics of the watermarked image.

\item
Robustness to common signal
attacks such as filtering and lossy compression.

\item
Lossless retrieval of embedded watermarks.

\item
User-friendly graphical interface for easy insertion and extraction of watermarks.
\end{enumerate}

\begin{figure*}[htbp]
\centering
\includegraphics[width=0.75\textwidth] {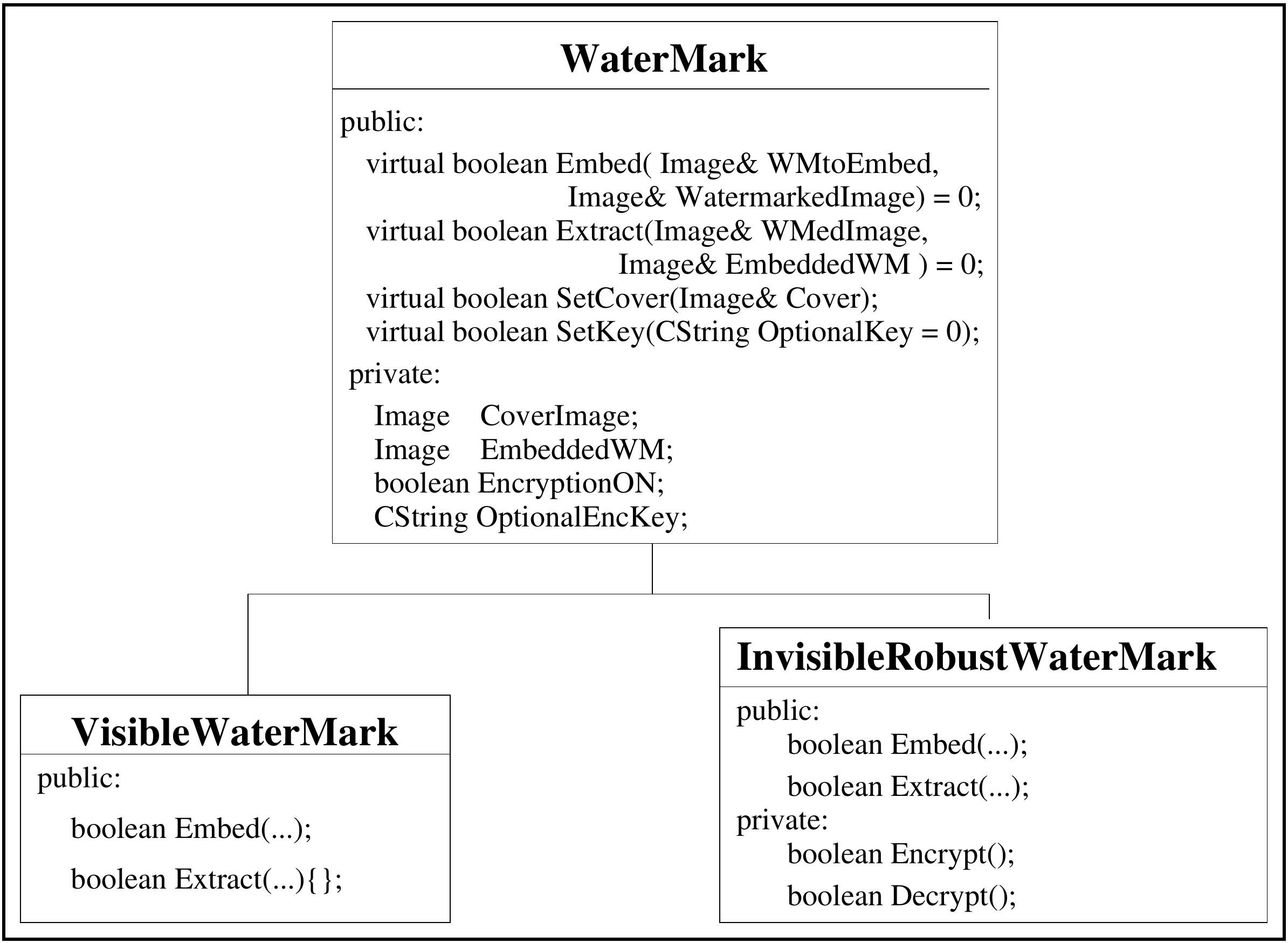}
\caption{Class hierarchy in ISWAR (default constructors and
destructors are not explicitly shown in diagram for brevity.)}
\label{fig:ISWAR-ClassHierarchy}
\end{figure*}

\begin{figure*}[htbp]
\centering
\subfigure[Visible watermark embedding.]
{\label{fig:VisibleWMScenario}
\includegraphics[height=4.0cm, width=0.48\textwidth] {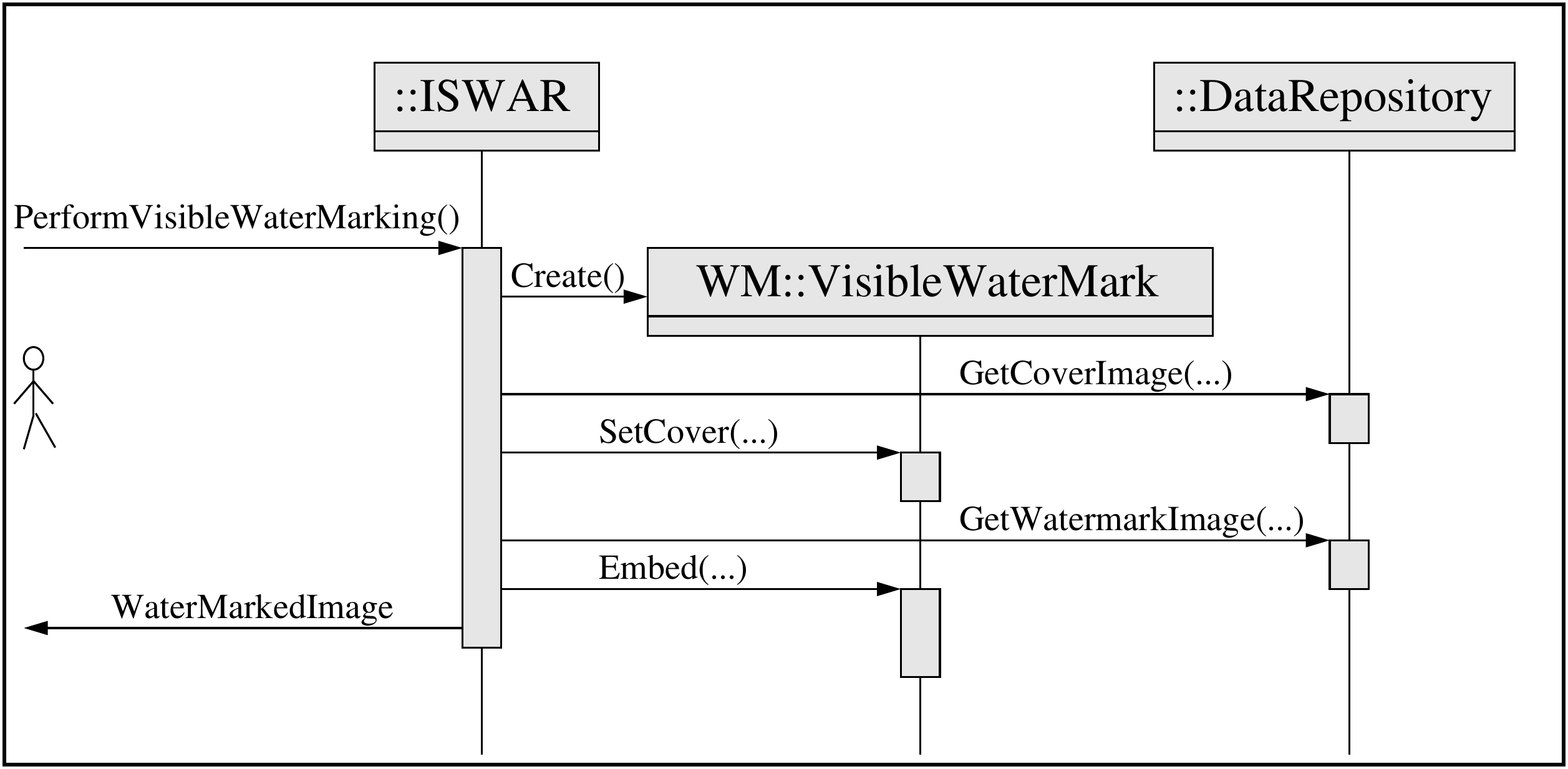}}
\subfigure[Invisible watermark embedding.]
{\label{fig:InvisibleWMScenario}
\includegraphics[height=4.0cm, width=0.48\textwidth] {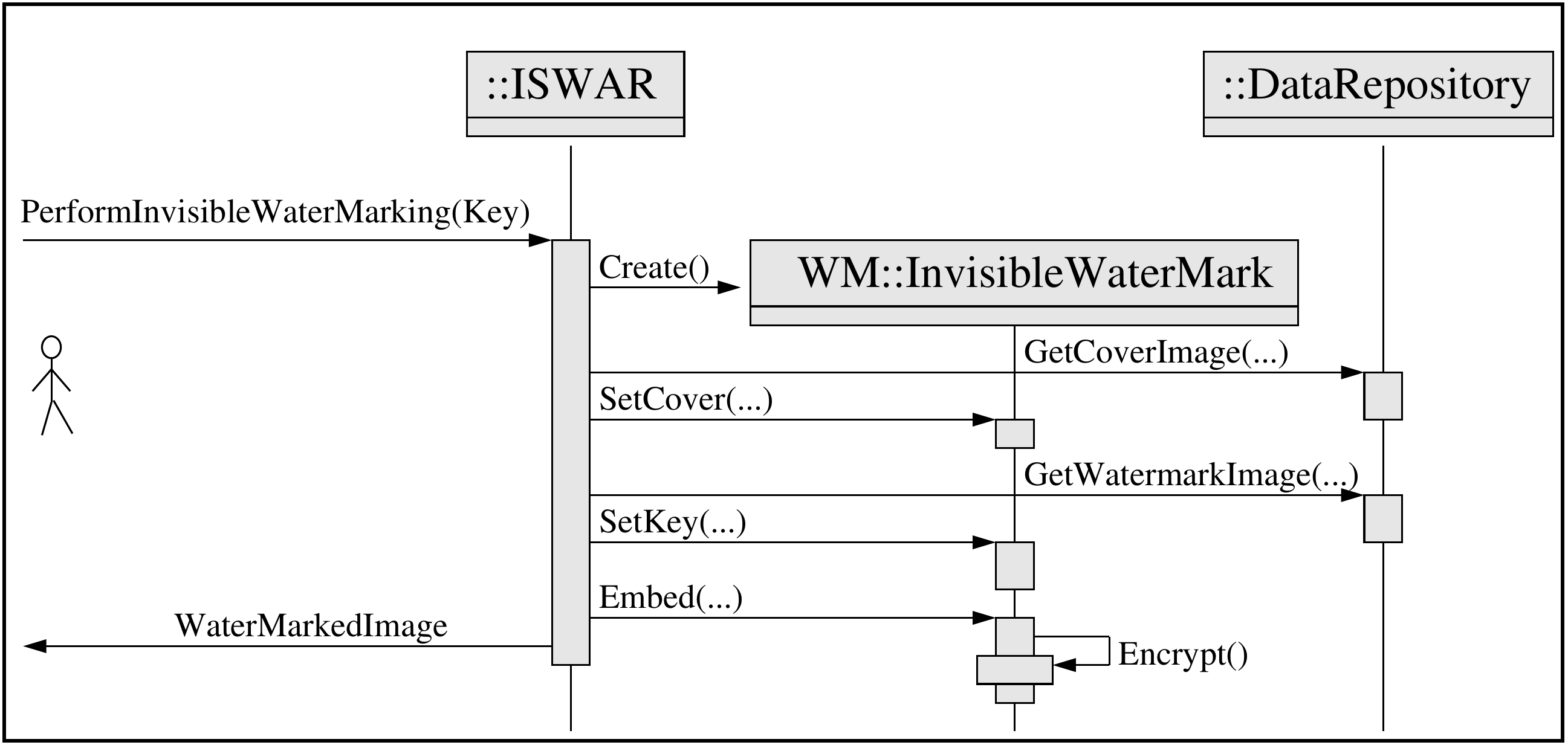}}
\subfigure[Invisible watermark extraction and authentication. The dotted arrows indicate optional operations required for non-blind extraction.]
{\label{fig:WMExtractionScenario}
\includegraphics[height=4.0cm, width=0.48\textwidth] {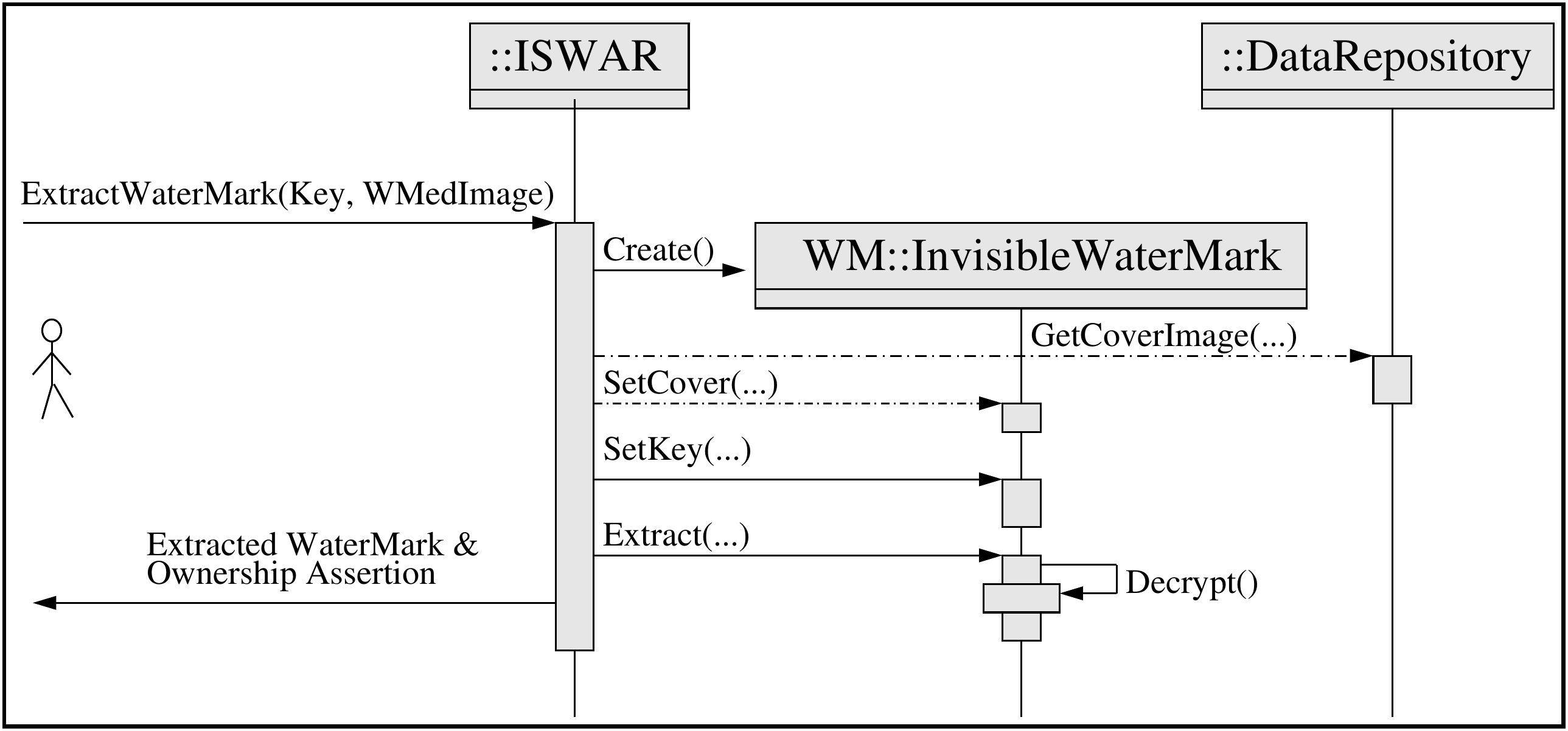}}
\caption{Sequence diagrams for different operations.}
\end{figure*}

The first requirement above translates into an
object-oriented (OO) architecture with a simple class hierarchy
depicted in Figure~\ref{fig:ISWAR-ClassHierarchy} for ISWAR.
In the OO analysis for ISWAR, the following three main scenarios are
considered: (1) Visible Watermarking, (2) Invisible Watermarking with
encryption, and (3) Extraction of encrypted invisible watermark. The
sequence diagrams corresponding to these 3 scenarios are presented
in Figures~\ref{fig:VisibleWMScenario}, \ref{fig:InvisibleWMScenario},
and \ref{fig:WMExtractionScenario}, respectively. The OO
architecture facilitates easy handling of new scenarios and
extensibility of the system with new types of watermarks and
watermarking or encryption algorithms.

The above analysis leads to the design of the algorithms for the
``Embed'' method that needs to be executed polymorphically for the
two types (\emph{i. e.,} visible and invisible) watermarks. The crux
of our design is in the two algorithms that we originally developed
for visible and invisible watermarking of monochrome images and
extended later for color images. The algorithms are presented in the
following sections. The watermark extraction method for images
embedded with invisible watermarks uses a process that is an inverse
of our invisible watermarking algorithm. It is described in a
subsequent section.

\section{Image Adaptable Visible Watermarking in ISWAR}
\label{VisibleWatermarkingAlgorithm}

Considering the robustness and data
reduction requirements of ISWAR, this paper focusses on the area of
transform domain (particularly the Discrete Cosine Transform, DCT, as it is omnipresent
in major compression standards) watermarking for the design of
our visible watermarking algorithm. For
ISWAR, this paper introduces a novel approach to visible watermarking
which is  image-adaptable and robust.
In this algorithm, unlike in other methods, the
classification of  image blocks into 8 perceptual classes for the
purpose of adaptable embedding is not performed. Rather, the parameters
involved in the watermark embedding for individual blocks are
computed using a mathematical model explicitly introduced for this purpose.
This model, based on a sensitivity analysis of the Human Visual System (HVS),
utilizes the statistics of the cover and watermark image blocks to calculate, for
each block, embedding parameter values that do not distort the cover image.

\subsection{Visible Watermark Embedding}

\begin{figure*}[htbp]
\centering
\includegraphics[width=12.0cm] {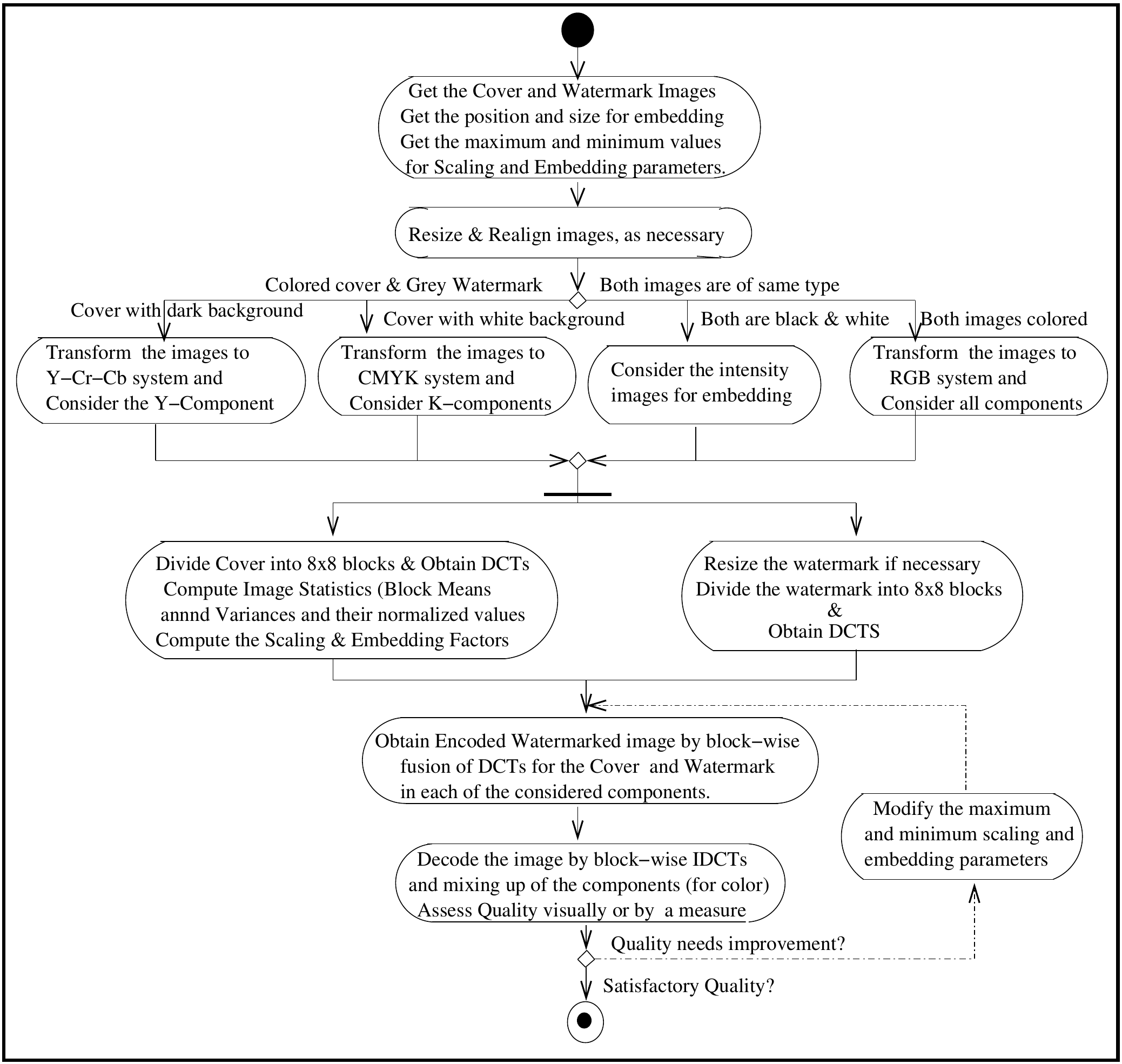}
\caption{Activity diagram for visible watermark embedding method.}
\label{fig:VisibleWMInsertionActivity}
\end{figure*}

The activity diagram for the visible watermark embedding process is shown
in Figure~\ref{fig:VisibleWMInsertionActivity}. Essential
inputs to the process are the cover and watermark image objects
including their sizes as well as the watermark's target size
(i.e., after embedding) and position for embedding. These are
provided through the ``Embed'' method of the watermark class. The
process can also be provisioned with four optional inputs, the
maximum and minimum values of the scaling and embedding parameters
($\alpha_{max}$, $\alpha_{min}$, $\beta_{max}$, and $\beta_{min}$)
that determine the relative proportions of the cover and watermark
contents in the embedding process. If these values are not
provided, pre-configured default values (presented in the
implementation section) are used by the system.

The algorithm assumes that the target size of the watermark
(i.e., the size after embedding) is equal to, or smaller than
than that of the cover image in both dimensions. Hence if
the specified target size of the watermark violates this stipulation,
the size should be properly reduced to bring in the watermark within
the bounds of the cover. Resizing is also required if the target
size is different from the original size of the watermark. Further,
since the algorithm works on $8 \times 8$ size image blocks, the
images need to be extended using standard image extrapolation
techniques, as necessary, to facilitate division of the original
images into an integral number of $8 \times 8$ blocks. The first step
of the algorithm is proper resizing of the images. Apart from the
image sizes, the watermark-position parameter may also
require adjustment. In the algorithm, this parameter is specified as
the number of the cover image block (in a system of successive
numbering with raster scanning) to which the watermark image is
aligned. Alignment involves positioning of the top-left corner of
the watermark image and the same corner of the specified block of the
image. However, if the specified position of the watermark is such
that a part of it falls outside the cover, the position is to be
readjusted to bring in the watermark totally into the range of the
cover. Thus, resizing and realignment are the operations in the
initial step of the algorithm. The next step is to select the
suitable components of the image for the embedding operation. In
case of grey level images, there is no choice; the algorithm has only the
intensity component. Similarly, when both cover and
watermark are colored, all three (R, G and
B) components of the images are considered and component by component
embedding is performed. In the case of colored cover and grey watermark,
it is advantageous to embed the watermark in the Y-component
(i.e., the luminance component in the Y-Cr-Cb coordinate system)
of the cover. More aesthetic
watermarked images may be obtained for grey watermarks and colored
covers with predominantly white background by using the K-components
of the CMYK color map of the cover \cite{TopkaraEtAlLecNotesCompSc2005}.
Thus, the characteristics of the
cover and watermark images determine the image coordinate space for
embedding. These characteristics may be assessed using a simple
analysis of the image pixels and a decision regarding the image
components required for the embedding operation may be taken as
shown in Figure~\ref{fig:VisibleWMInsertionActivity}.

Once the component(s) is (are) are obtained, each component of the
cover and the watermark is divided into blocks of size $8 \times 8$.
For each block of the cover and the watermark images, the DCT is
obtained, if possible, by parallel processing. Let us denote by
$c_n$, $w_n$, and $c'_n$ the $n$-th blocks, in the DCT domain, of
the cover, the watermark, and the watermarked images, respectively.
We also denote by $c_{ij}(n)$, $w_{ij}(n)$, and $c'_{ij}(n)$, the
$(i,j)$-th DCT coefficient of the $n$-th block of the respective
images; when $i=j=0$, the corresponding coefficient is DC.

The next step of the algorithm is to gather image statistics for the
cover. This information is useful for computation, as shown in the
following subsection, of the scaling and embedding parameters
necessary for block-by-block fusion of the cover and the watermark
images. The statistics comprise of the block mean intensities and
log variances (AC DCT coefficients) of individual blocks. The mean
values of individual blocks are given by the DC DCT coefficients of
the blocks. These values are scaled in the range $[0.1-1.0]$ and
normalized block means ($\mu'_n$ s) are obtained using the
following expression:
\begin{equation}
\mu'_n = 0.1 + 0.9 \left( \frac { c_{00}(n) -  {c_{00}}_{min} } {
{c_{00}}_{max} -  {c_{00}}_{min} } \right),
\end{equation}
where, ${c_{00}}_{max}$ and ${c_{00}}_{min}$ are the maximum and
minimum values, respectively, among the DC DCT coefficients of
different blocks of the cover. Assuming the number of blocks in the
image is $N$, the normalized mean gray value of the image $I$ is
calculated using the following expression:
\begin{equation}
\mu' = \frac{1}{N} \sum^N_{n=1} \mu'_n.
\end{equation}
The log variances for the AC DCT coefficients of individual blocks
are obtained using the formula:
\begin{equation}
\sigma_n = \log_e \left( \frac{1}{63}\sum_i \sum_{j|i+j\neq 0} \left(
c_{ij}(n) - \mu_{n}^{ac} \right)^2 \right),
\end{equation}
where $\mu_{n}^{ac}$ is the average of the AC DCT terms of the
$n$-th block obtained by the following expression:
\begin{equation}
\mu_{n}^{ac} = \frac{1}{63}\sum_i \sum_{j|i+j\neq 0} c_{ij}(n).
\end{equation}
These log variances are scaled in the range $[0.1-1.0]$ and
normalized block log variances are obtained by the following
expression:
\begin{equation}
\sigma'_n = 0.1 + 0.9 \left( \frac{ \sigma_n - \sigma_{min} } {
\sigma_{max} - \sigma_{min} } \right).
\end{equation}
Here, $\sigma_{max}$ and $\sigma_{max}$ are the maximum and minimum
values, respectively, of the log variances for different blocks of
the cover. In addition to gathering these image statistics,
further image analysis to identify edge blocks in the cover is
required for the purpose of computing the scaling and embedding
parameters as indicated in the following subsection.

The final step in visible watermarking is the embedding operation
itself. For this, the watermark is positioned on the cover at the
specified position (after all adjustments in the initial step of the
algorithm). For portions of the cover image not covered by such a
superposition of the watermark, the watermark pixel values (hence the
corresponding block DCT coefficients) are assumed to be zero.  Now,
the watermarked image may be conceived as the one obtained by fusing
the original watermark, so augmented with zero pixel values, into
the cover. The DCT domain block-wise fusion operation may now be defined by the following equation:
\begin{equation}\label{Eq:Visible-WM-Embed}
c'_{ij}(n) = \alpha_n \times c_{ij}(n) + \beta_n \times w_{ij}(n).
\end{equation}
$\alpha_n$ and $\beta_n$ are the scaling and embedding parameters
that need to be chosen on a per block basis in such a way as to
yield a watermarked image that is appealing to to the HVS
\cite{GranrathProceedingsIEEE1981}. In the following subsection, we
develop mathematical formulae for these parameters based on the
information available in the literature on visual perception
regarding the contextual sensitivity of HVS to distortions. By
juxtaposition of the $c'_{ij}(n)$ blocks obtained as above, we obtain
the encoded image $I'$. The watermarked image in spatial form may be
obtained from $I'$ by block-wise decoding the individual
$c'_{ij}(n)$s with IDCTs (Inverse DCTs). Hence, an important
consideration in the choice of the above scaling and embedding
parameters is to contain, to imperceptible levels, the degradation
in  the resultant image due to blocking (or tiling) effects in the
inverse transform process and other distortions. Another
consideration is to make the watermark more robust in the sense that
removal or destruction of the watermark will result in a degraded image
that cannot be reused. The watermarked image is expected to be of
good quality for the proper choice of the scaling and embedding
parameters. The visual quality may be assessed by either visual
inspection or any application-specific quality measures. The optional
feedback loop (dashed line) in the activity diagram suggests that
this information regarding image quality may be used to tone up the
image quality by tuning either manually or automatically the
parameters $\alpha_{max}$, $\alpha_{min}$, $\beta_{max}$, and
$\beta_{min}$ and, through them, the computations of the $\alpha_n$
and $\beta_n$ values for different blocks.

\subsection{Determining Scaling and Embedding Factors - A Mathematical Model}

Our mathematical model for determination of  the scaling and
embedding factors, $\alpha_n$ and $\beta_n$, is based on the
following psycho-visual considerations that help to preserve the
perceptual quality of the watermarked images despite distortions:
\begin{itemize}
\item
The edge blocks should be least altered to avoid significant
distortion of the image and hence the relative intensity of the
watermark should be low in any block containing edges of the cover
image. This, in turn,  means that $\alpha_n$ and $\beta_n$ should be
set to values close to $\alpha_{max}$ and $\beta_{min}$ (the
user-configured maximum and minimum value of the scaling and the
embedding factors), respectively.

\item
The distortion visibility is low when the background has strong
texture. The higher the texture of a block, the lower the variance of
its AC DCT coefficients due to more evenly distributed energy among
these coefficients. This suggests that proportionately larger
information content of the watermark can be fused into those blocks
without adversely affecting the distortion visibility. Hence, we
assume that a choice of $\alpha_n$ directly proportional to the
variance $\sigma_n$ of the AC DCT coefficients of $c_{ij}(n)$, and a
choice of $\beta_n$ inversely proportional to the same would be
ideal.

\item
The blocks with mid-intensity are more sensitive to noise than those
of low and high intensities. This suggests a bell-shaped Gaussian
distribution for $\alpha_n$ values; they should increase with the mean
intensity values ($\mu_n$s) of their respective blocks up to a mid
intensity value (say, the mean intensity $\mu$ of the whole cover
image) and then start decreasing with the block mean intensities. As
usual, the variation of the $\beta_n$s should be exactly the
reverse. The $\mu_n$s are given by the DC DCT coefficients of the
corresponding blocks.
\end{itemize}

\begin{figure*}[htbp]
\centering
\subfigure[Variation of $\alpha_n$
with $\sigma_n$] {\label{fig:AlphanVsSigmanVariation}
\includegraphics[width=6.8cm] {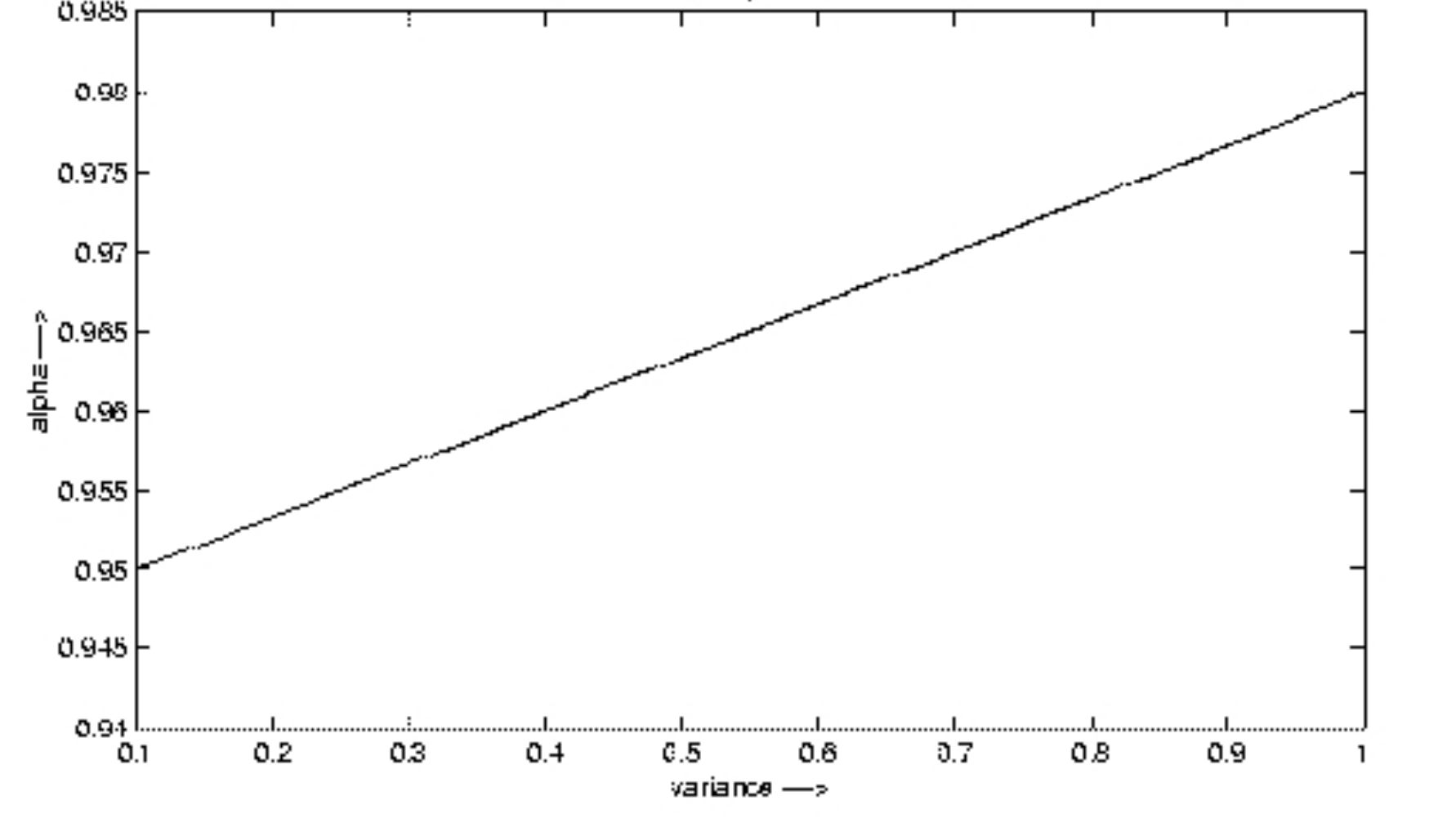} }
\subfigure[Variation of $\beta_n$ with $\sigma_n$]  {
\label{fig:BetaVsSigmanVariation}
\includegraphics[width=6.8cm] {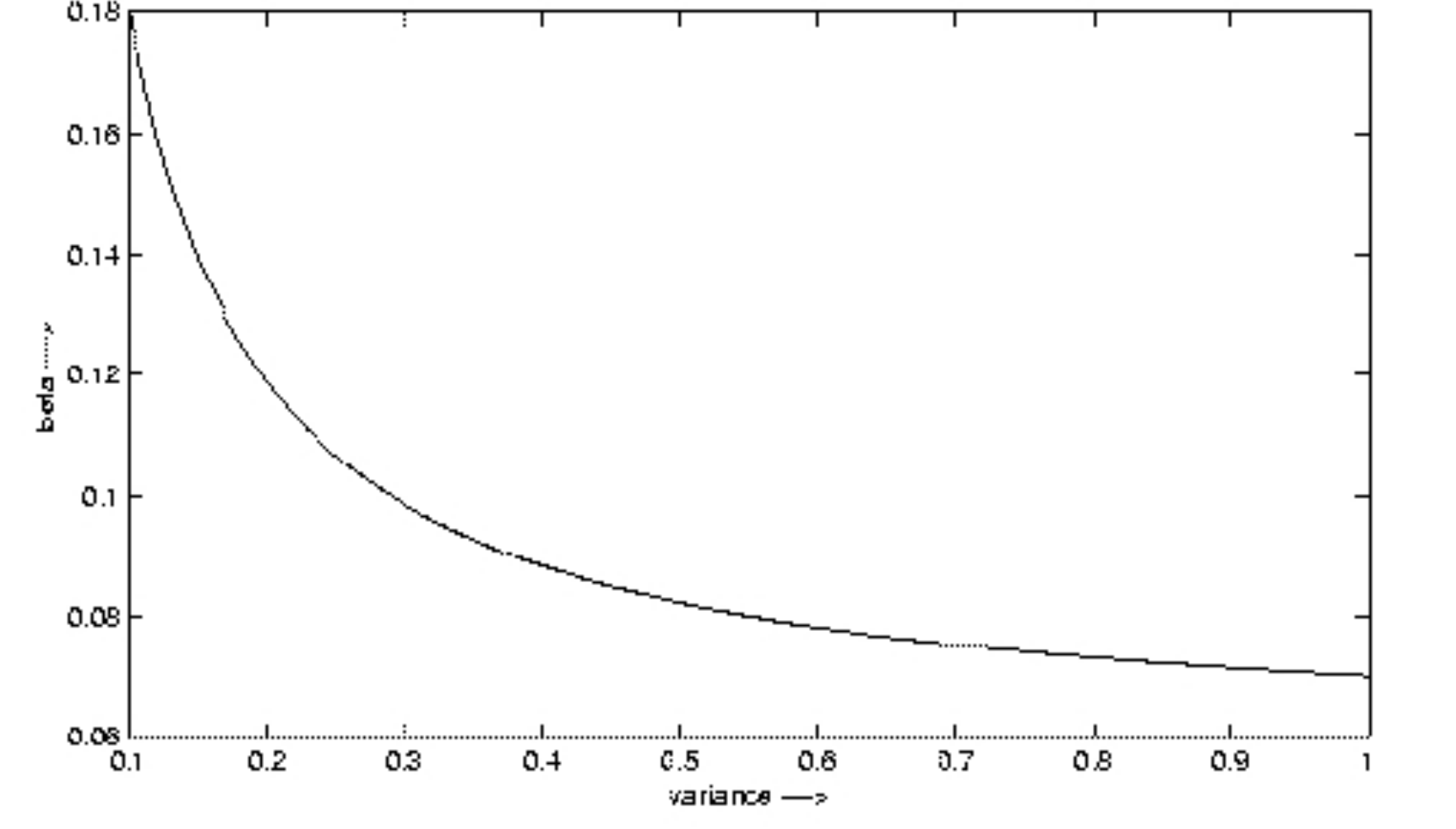} }
\subfigure[Variation of $\alpha_n$ with $\mu_n$]  {
\label{fig:AlphanVsMunVariation}
\includegraphics[width=6.8cm] {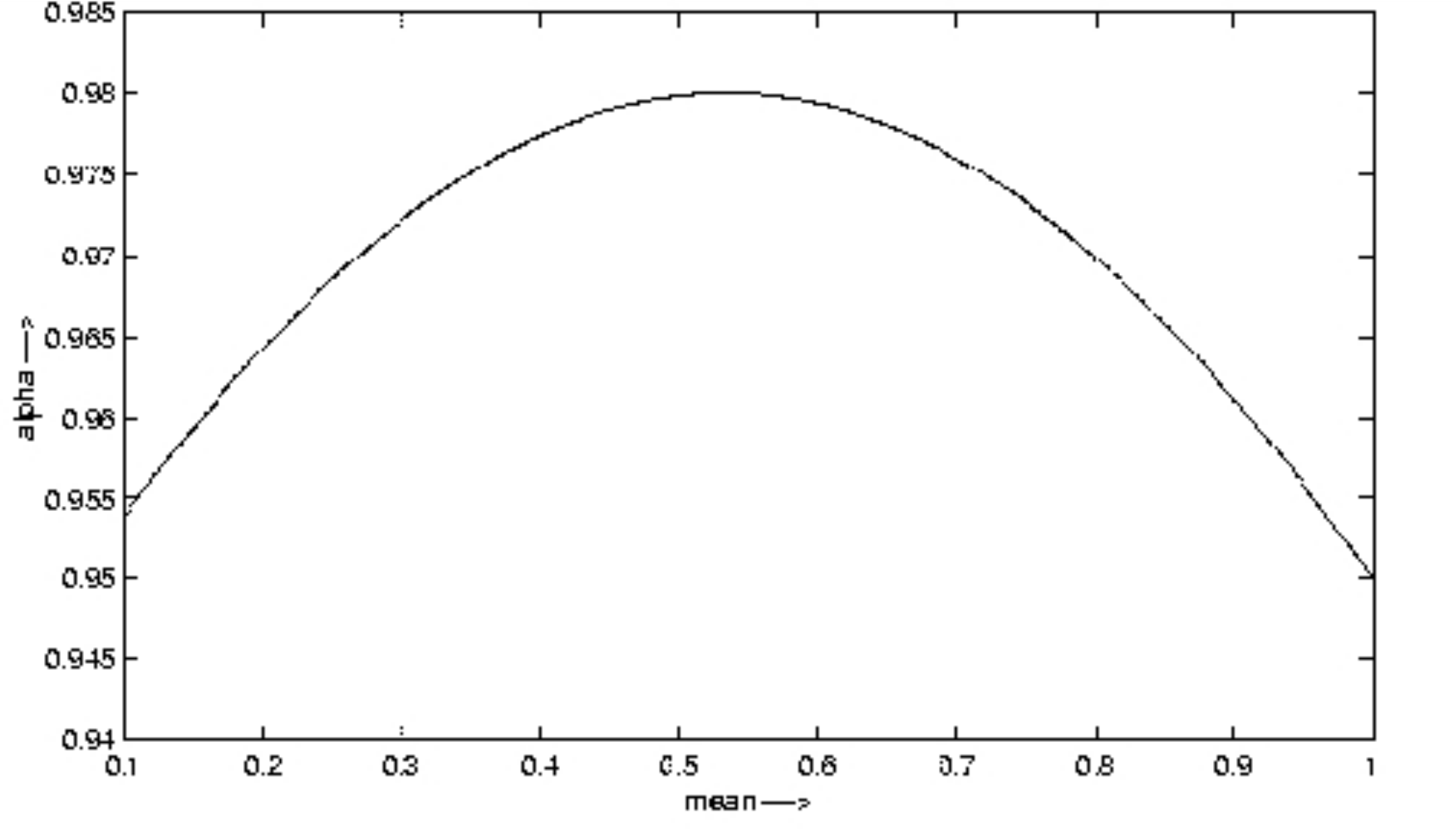} }
\subfigure[Variation of $\beta_n$ with $\mu_n$]  {
\label{fig:BetanVsMunVariation}
\includegraphics[width=6.8cm] {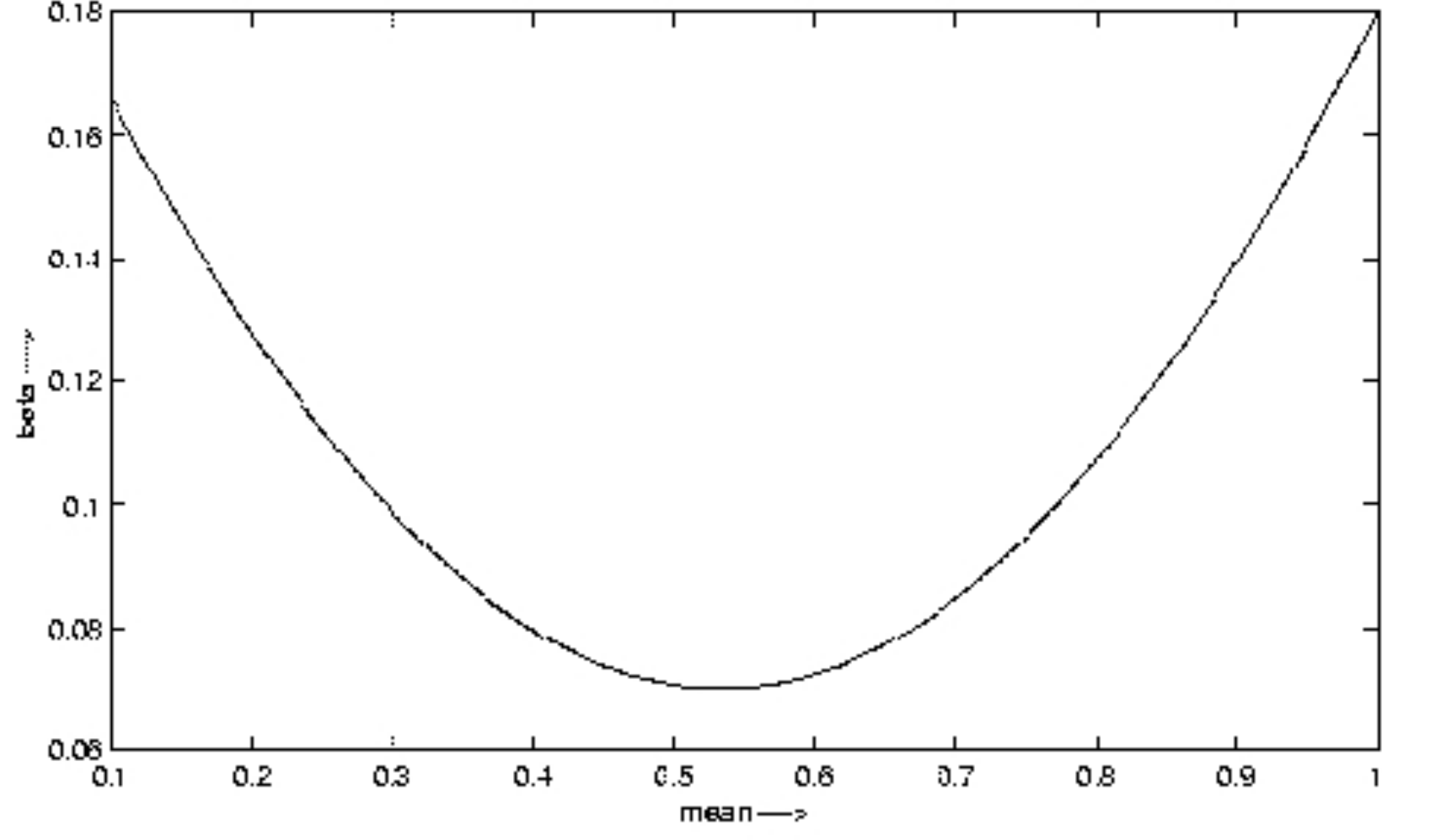} }
\caption{Variation of scaling and embedding factors with respect to
cover image statistics.} \label{fig:Alpah_Beta_Variations}
\end{figure*}

Figure~\ref{fig:Alpah_Beta_Variations} depicts the variation of the
scaling and embedding factors based on these theoretical
considerations. Mathematical formulation of these relationships
yields the following equations for the scaling and embedding
parameters:
\begin{eqnarray}
\mbox{\hspace{-0.4cm}}\alpha_n & = & \left\{
\begin{array}{ll}
\alpha_{max} & \mbox{For Edge Blocks} \\
\alpha_{min} + \left(\alpha_{max} - \alpha_{min}\right) \\
\mbox{\hspace{0.3cm}} \times \left[\sigma'_n \mbox{\hspace{0.1cm}} \exp\{
-(\mu'_n-\mu')^2 \} \right] & \mbox{For Other Blocks}
\end{array} \right. \\
\mbox{\hspace{-0.4cm}}\beta_n & = & \left\{
\begin{array}{ll}
\beta_{min} & \mbox{\hspace{-0.5cm} Edge Blocks} \\
\beta_{min} + \left(\beta_{max} - \beta_{min}\right) \\
\mbox{\hspace{0.01cm}} \times \left[ \left( \dfrac{1}{\sigma'_n} \right)
\mbox{\hspace{0.01cm}} \left\{ 1 - \exp \left( -(\mu'_n-\mu')^2
\right) \right\} \right]& \mbox{\hspace{-0.2cm} Other Blocks}
\end{array} \right.
\end{eqnarray}
The parameters $\alpha_{max}$, $\alpha_{min}$, $\beta_{max}$ and
$\beta_{min}$ could be specified by the user. If they are not
specified, pre-defined default values may be used. On
experimentation with a large number of images, the above expressions
have been found to conform to the patterns of the theoretical
relationship discussed earlier.

\section{Secure Invisible Watermarking in ISWAR}
\label{InvisibleWatermarkingAlgorithm}

Because of the data compression and fusion robustness aspects of the
transform domain methods, we focus here also on the transform domain
approach to invisible watermarking. For this initial version of ISWAR,
a simple but effective invisible watermarking algorithm is
introduced that uses a binary watermark which is described in the following subsections.

\subsection{Invisible Watermark Embedding}

\begin{figure*}[t]
\centering
\includegraphics[width=11.5cm] {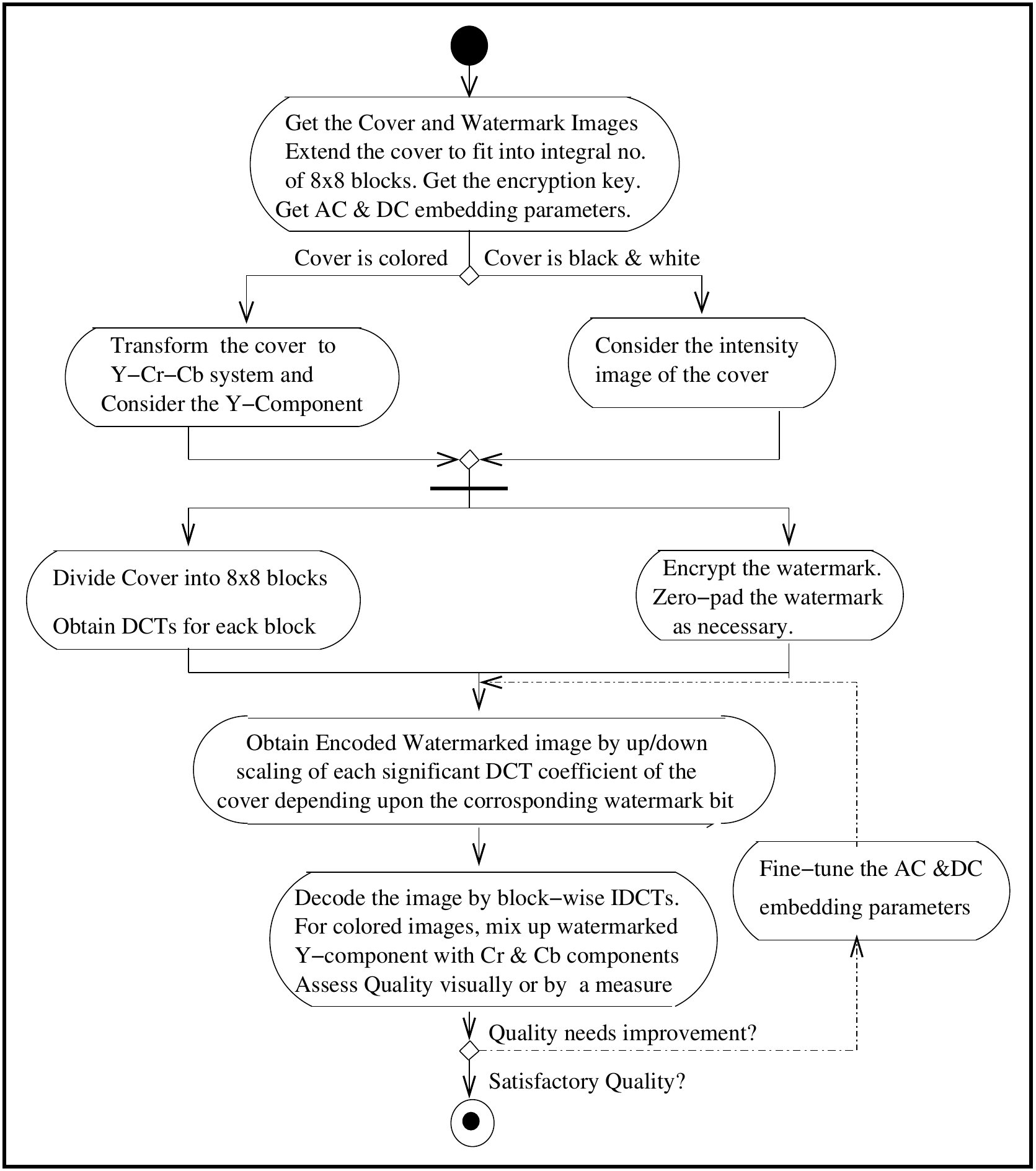}
\caption{Activity diagram for invisible watermark embedding method.} \label{fig:FlowOfInvisibleInsertion}
\end{figure*}

Figure~\ref{fig:FlowOfInvisibleInsertion} shows the UML activity
diagram for our invisible watermark insertion process.  To achieve
two levels of protection as indicated earlier, the algorithm first
encrypts the  watermark  and then fuses it into the intensity image
of the cover in the case of a black and white cover, and into the
Y-component (in the Y-Cr-Cb coordinate system) of a colored cover.
In the sequel, the relevant component of the cover is referred as
$I$. Decomposition of the image to obtain  the required component is
done in the preprocessing stage of the algorithm. Encryption and
hashing of the binary watermark using a user-supplied key is also
performed in this preprocessing step. Further, just as in the
visible watermarking algorithm, any image extension necessary to
facilitate the division of the image into integral number of blocks
is performed in this stage.

After preprocessing, the cover image $I$ is divided  into
$8 \times 8$ blocks and each block is transformed into the DCT domain.
Let us denote the ``$(i,j)$''-th DCT coefficient of the $k$-th block by
$c_{ij}(k)$. Assuming that the image has $M$ blocks overall, each
block can be numbered uniquely with a number in the range $[1,M_{block}]$
based on its position in the raster scanning of the image.
$M_{block}$ is given by $(M \times N)/64$,
where $M$ is the number of image pixels row-wise, and $N$, the
number of pixels column-wise.
The number of DCT components to be considered for obtaining good quality
watermarked images is a tradeoff of quality and robustness.
In this algorithm, the DC component
$c_{00}$ and the three 3 low frequency components $c_{01}$,$c_{10}$
and $c_{11}$ are considered. In this case, the size of the encoded and hashed
watermark should be such that it can be partitioned into the same
number of blocks as the cover, but with a block size of $2 \times
2$. It cannot be bigger, but, if it is smaller, it can be padded
with zeros. Using the same notation as before,
the watermark's binary value at position $(i,j)$ in block $k$
is denoted as $w_{ij}(k)$. This watermark is embedded
in the cover image using the following expression:
\begin{equation}
\begin{array}{ll}
\forall i,j, \ {\mbox and}\  k, \\
c'_{ij}(k) = \left \{
\begin{array}{ll}
c_{ij}(k) (1 + \alpha_{ij})\ \  if\ \  w_{ij}(k) = 1, \\
c_{ij} (k) (1 - \alpha_{ij})\ \  if\ \  w_{ij}(k) = 0.
\end{array}\right .
\end{array}
\end{equation}

In this algorithm, the watermark is not always added
to the significant frequency components. Instead, the watermark
is added to some components and subtracted from others.
This satisfies the system requirement that a statistical  analysis of
watermarked image should not reveal the presence of an invisible
watermark. Another feature of the proposed method
is that two different scaling factors
($\alpha$s) are used for different frequency components:
$\alpha_{dc}$ for DC components and $\alpha_{ac}$ for AC
components. Thus, the algorithm has the
following: $\alpha_{00} = \alpha_{dc}$ and
$\alpha_{01} = \alpha_{10} = \alpha_{11} = \alpha_{ac}$. Since
choosing so many scaling factors (one for each frequency component)
is a problem by itself, the algorithm is confined to only two values
that may be so chosen so as not to degrade the quality of the
watermarked image. Image quality can be assessed either
qualitatively by visual inspection or quantitatively by measuring
its Peak Signal-to-Noise Ratio (PSNR).

In the case of black and white covers, the watermarked image $I'$ can be
obtained by performing block-wise IDCTs (Inverse Discrete
Transforms) on the coefficients modified as above. However, in the case
of colored images, we get only the Y-component of $I'$ by the above
process. This should be clubbed with the Cr and Cb components of the
cover to get $I'$. At this point, as in the visible watermarking
case, the algorithm has an optional step (dashed line)of assessing quality
of the watermarked image by either visual inspection or a
computational measure and fine-tuning of the parameters $\alpha_{ac}$
and $\alpha_{dc}$.

\subsection{Invisible Watermark Extraction and Authentication Method of ISWAR}

\begin{figure*}[t]
\centering
\includegraphics[width=11.0cm] {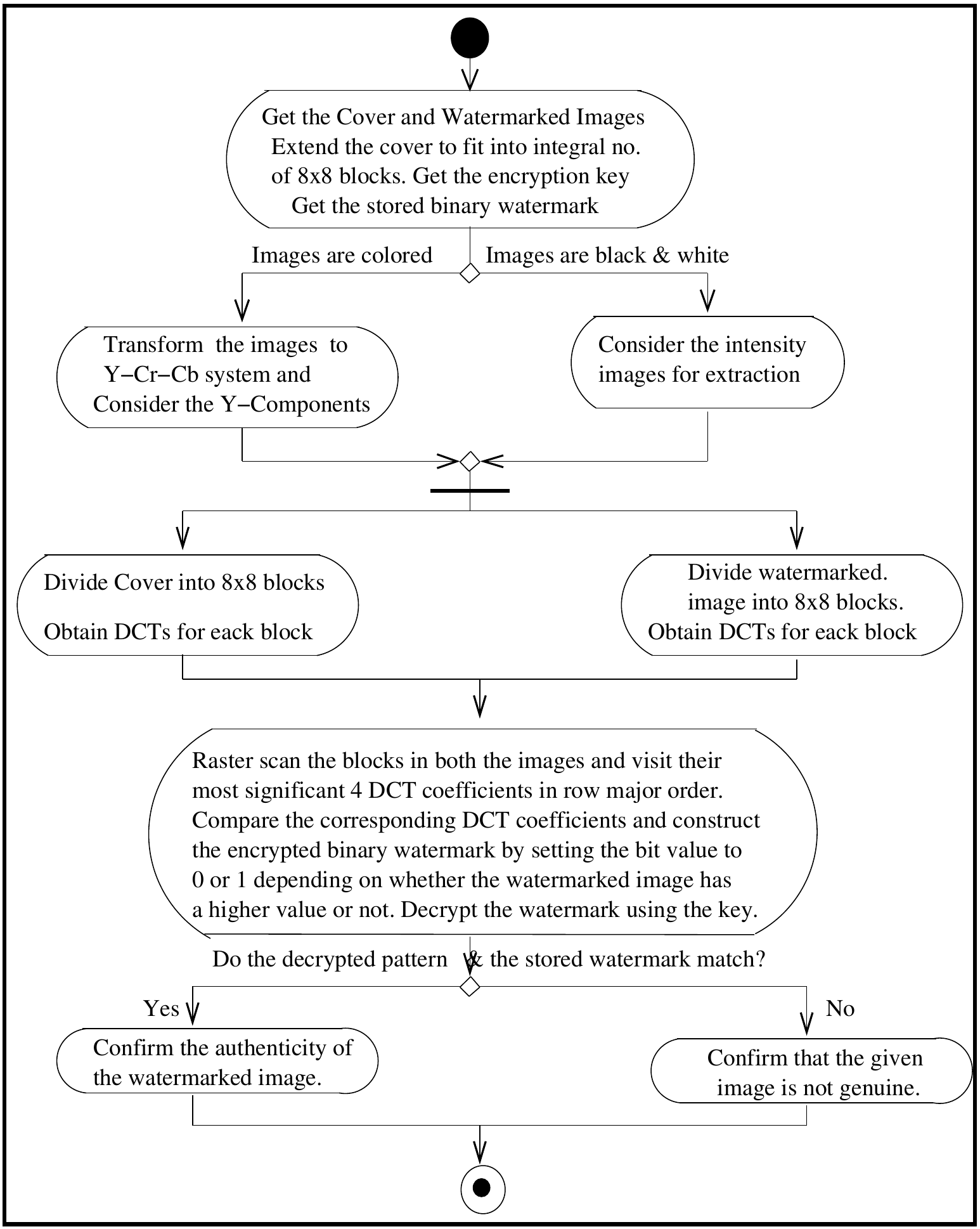}
\caption{Activity diagram for the invisible watermark extraction
and authentication method.} \label{fig:FlowOfInvisibleExtraction}
\end{figure*}

Figure~\ref{fig:FlowOfInvisibleExtraction} depicts the activity
diagram for the invisible watermark extraction process in ISWAR. The
extraction algorithm involves many steps which are reversals
of the insertion process. The algorithm first
gets the watermarked (possibly suspect) image $I'$ and the original cover image $I$.
The watermark image and key that is to be tested are also gathered.
The algorithm divides both $I'$ and $I$ into $8 \times 8$ blocks
as in the insertion process.
Both the cover and watermark images are
converted from RGB space to YCbCr representation if they are color.
DCT transformation is performed for the blocks from both images.
The algorithm at this point
compares corresponding image blocks.
If a DCT coefficient in a block of $I'$ is larger than the
corresponding coefficient in the original image block then the watermark
bit is 1, else it is 0.
The algorithm compares the extracted sequence with the binary watermark (encrypted
with the key) to make a decision whether the image is authentic or not.

\section{Implementation, Usage and Validation of ISWAR}
\label{sec:ImplementationUsageValidation}

ISWAR is implemented using the VC++.NET platform \cite{MohantyISCE2007}.
In order to be able to view an
image in the document window, it should be saved in one of the
standard image formats such as jpeg (.jpg), bit-map (.bmp), portable
network graphics (.png), or graphics interchange format (.gif).
The system defaults for
$\alpha_{max}$, $\alpha_{min}$, $\beta_{max}$ and $\beta_{min}$ are
set to 0.95, 0.98, 0.05 and 0.17, respectively. These values have
been found to yield good results and hence have been incorporated
into the system as defaults. For the same reason,
the system has been provisioned with the default values of 0.1 and 0.02 for
the parameters $\alpha_{ac}$ and $\alpha_{dc}$ used in the invisible
watermarking algorithm. For identifying edge blocks in the
visible watermarking case, the Sobel edge operator is used because
of its simplicity. It could be replaced by more sophisticated
operators in later versions of ISWAR. The encryption algorithm used in our secure
invisible watermarking was blow-fish. Any other encryption algorithm, e.g. Advanced Encryption Standard (AES) can be easily integrated in the future.
ISWAR has a user-friendly interface through which choice of the
algorithms and the specification of parameters can be performed.

\begin{figure*}[t]
\centering
\subfigure[Open an image.]
{\label{fig:Image_Open}
\includegraphics[height=5.2cm] {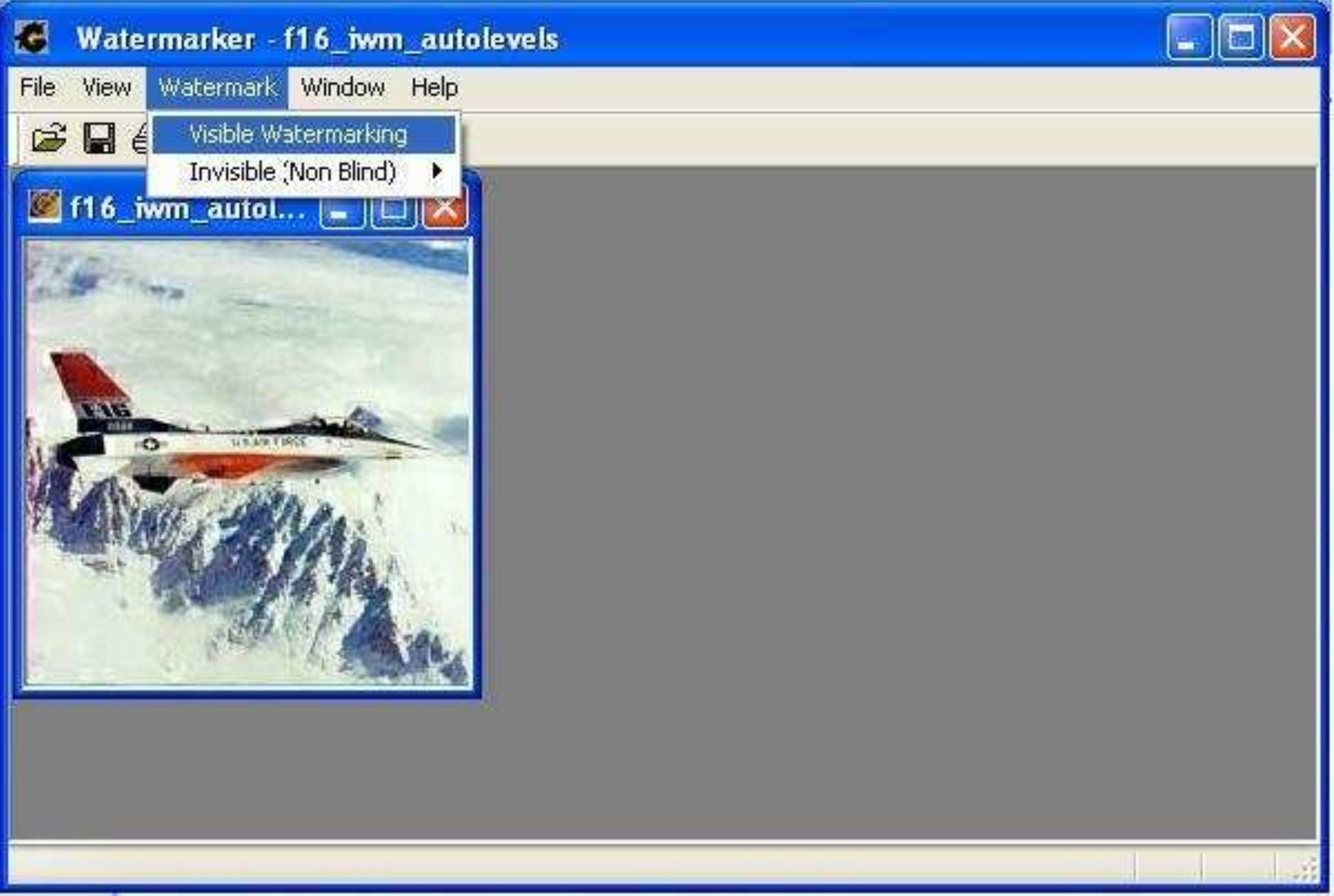} }
\subfigure[Select position, size and intensity.]  {
\label{fig:visible_parameters}
\includegraphics[height=5.2cm] {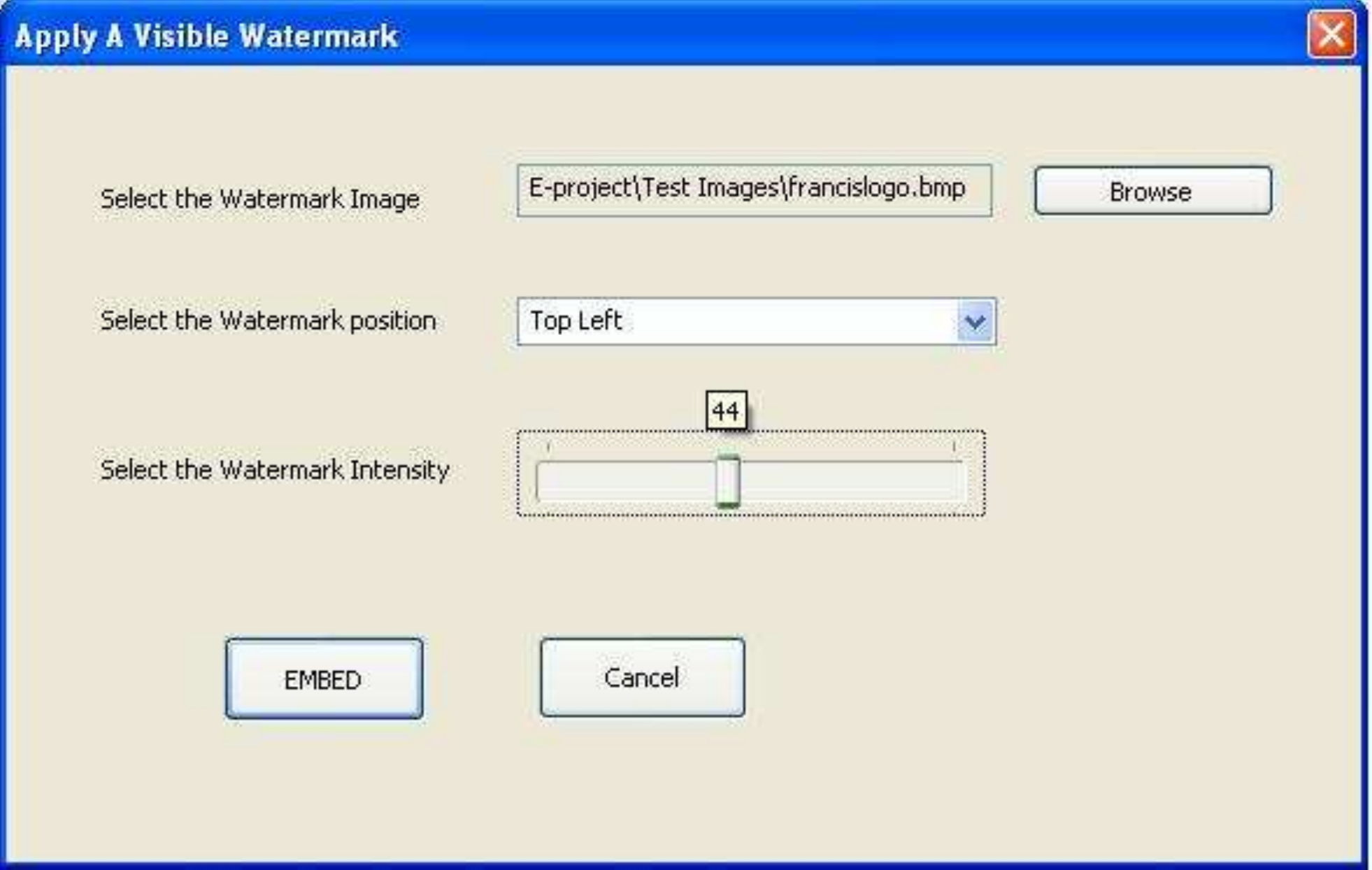} }
\subfigure[Insert the watermark.]  { \label{fig:visible_embed}
\includegraphics[height=5.2cm] {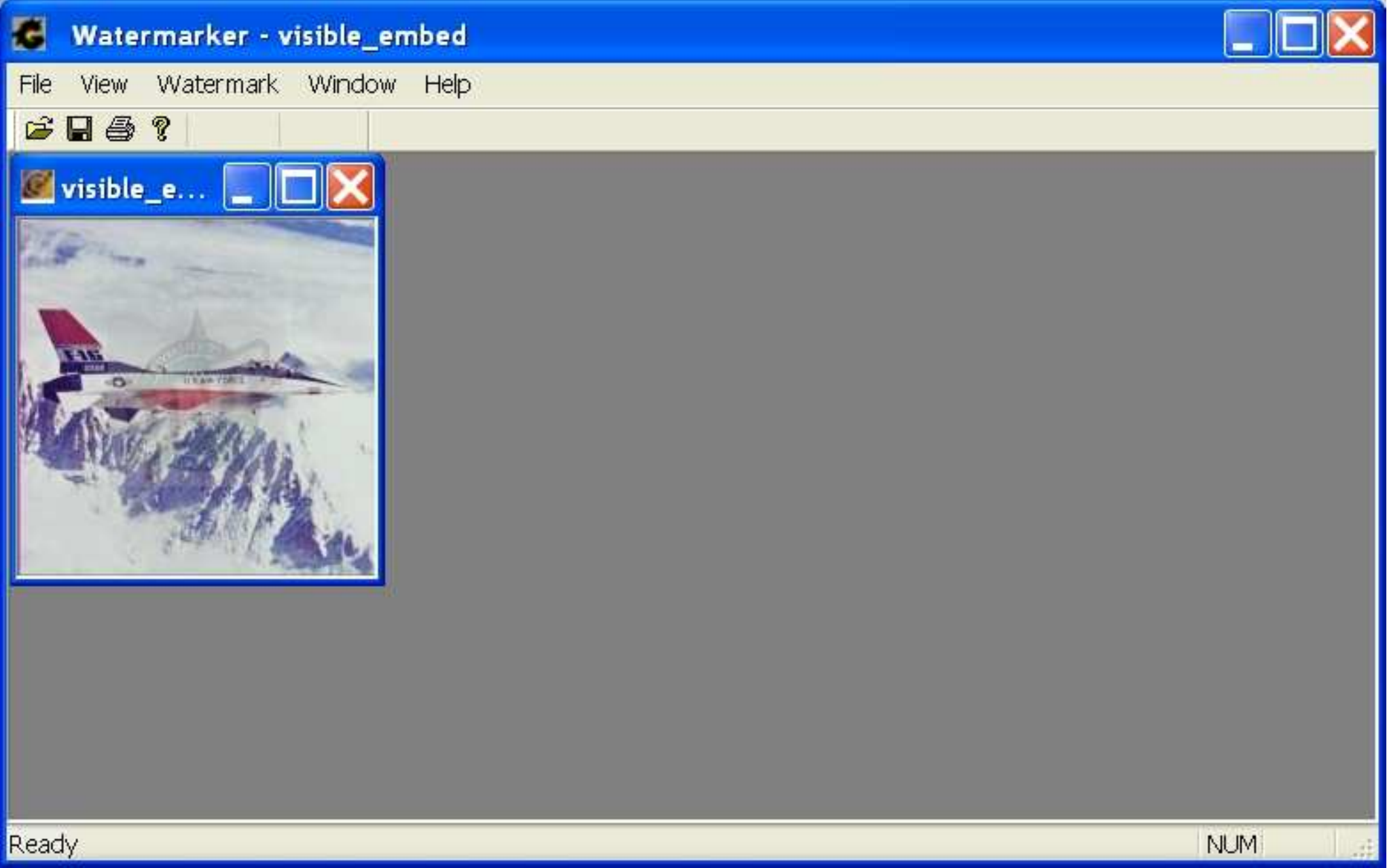} }
\caption{Performing visible watermarking using ISWAR.}
\label{fig:TestingVisibleInsertionAlgoInISWAR}
\end{figure*}

The visible watermark embedding can be initiated as shown in
Figure~\ref{fig:TestingVisibleInsertionAlgoInISWAR} using a file menu
for selecting and opening a cover image and watermark menu for
selecting the required operation.
Once the cover and watermark images are stored, visible watermarking
may be performed using the following steps, shown in Figure~\ref{fig:TestingVisibleInsertionAlgoInISWAR}:
\begin{itemize}
\item
Open the host (cover) image by ``Click File ---$>$ Open'' or the
Shortcut ``.''
\item
Select ``Apply Visible Watermark'' from the Watermark menu. This
will open up a new screen (dialog box) with the same name.
\item
Select the watermark file by either typing the path to
the file in the list box on the screen, or browsing using the
``Browse'' button. The size (height and width) of the watermark
image should be less that of the host image that was selected.
\item
Specify the watermark position using the ``Combo'' (combination)
box on the same screen. The 9 choices available for selection are as follows:
(1) Top Left, (2) Top Center, (3) Top Right, (4) Middle Left, (5) Middle
Center, (6) Middle Right, (7) Bottom Left, (8) Bottom Center, and
(9) Bottom Right. These positions are relative to the host image.
\item
Specify the Intensity of the watermark image to be embedded in
the host image using the slider control available on the screen. The
intensity values range from 10 to 100 and the selected value will be
displayed on the sliding bar.
\item
Click the ``Embed'' button. This button click will generate the
event trigger to run the algorithm to insert the visible watermark.
After the algorithm is executed, the dialog box for Visible
Watermark will close and the watermarked image will be opened in the
document window.
\end{itemize}

\begin{figure*}[t]
\centering
\subfigure[Open an image.]
{\label{fig:Image_open_iem}
\includegraphics[height=5.0cm] {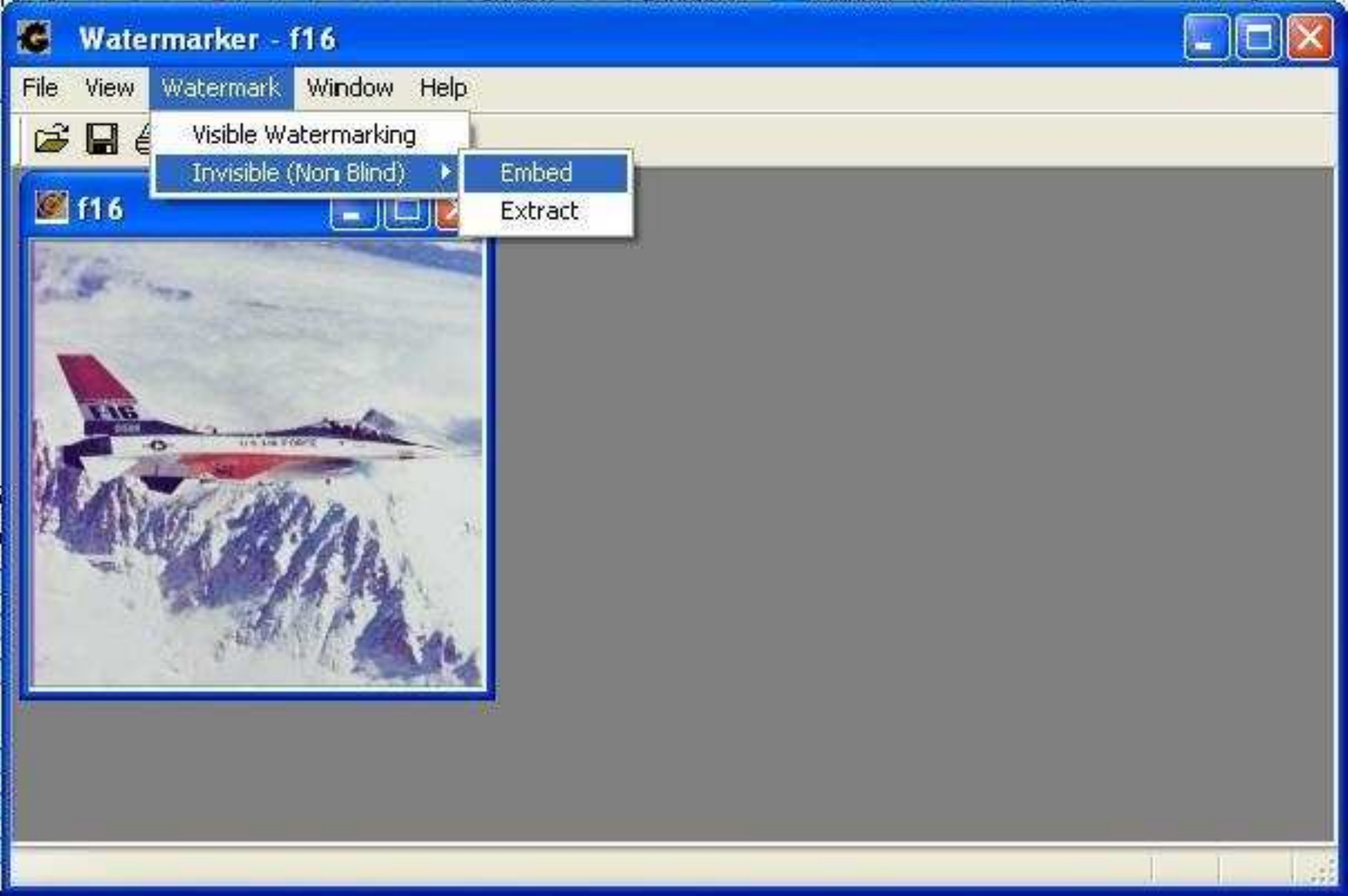} }
\subfigure[Enter watermarking key.]  {
\label{fig:invisible_embed_parameters}
\includegraphics[height=5.0cm] {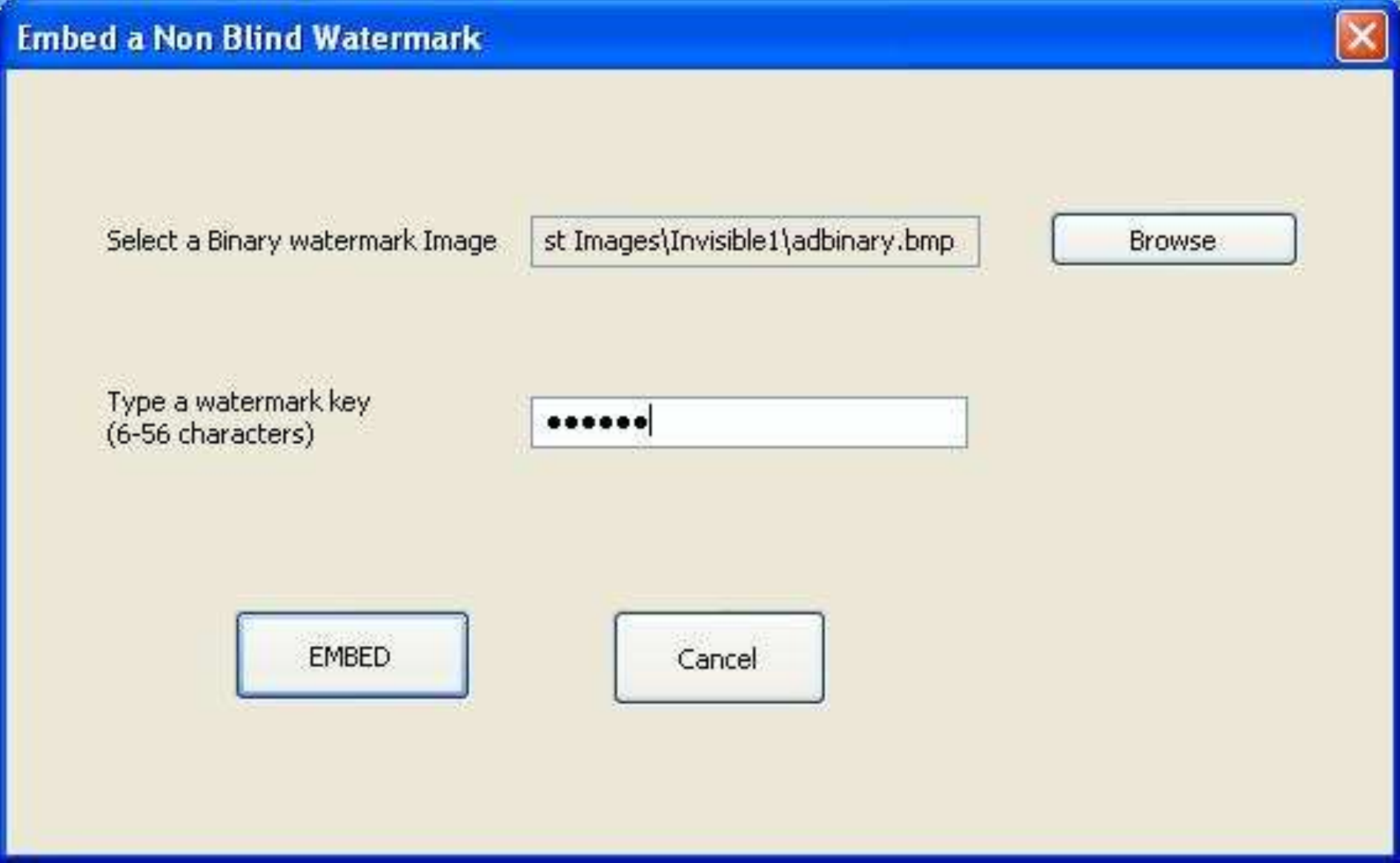} }
\subfigure[Insert the watermark.]  { \label{fig:invisible_embed}
\includegraphics[height=5.0cm] {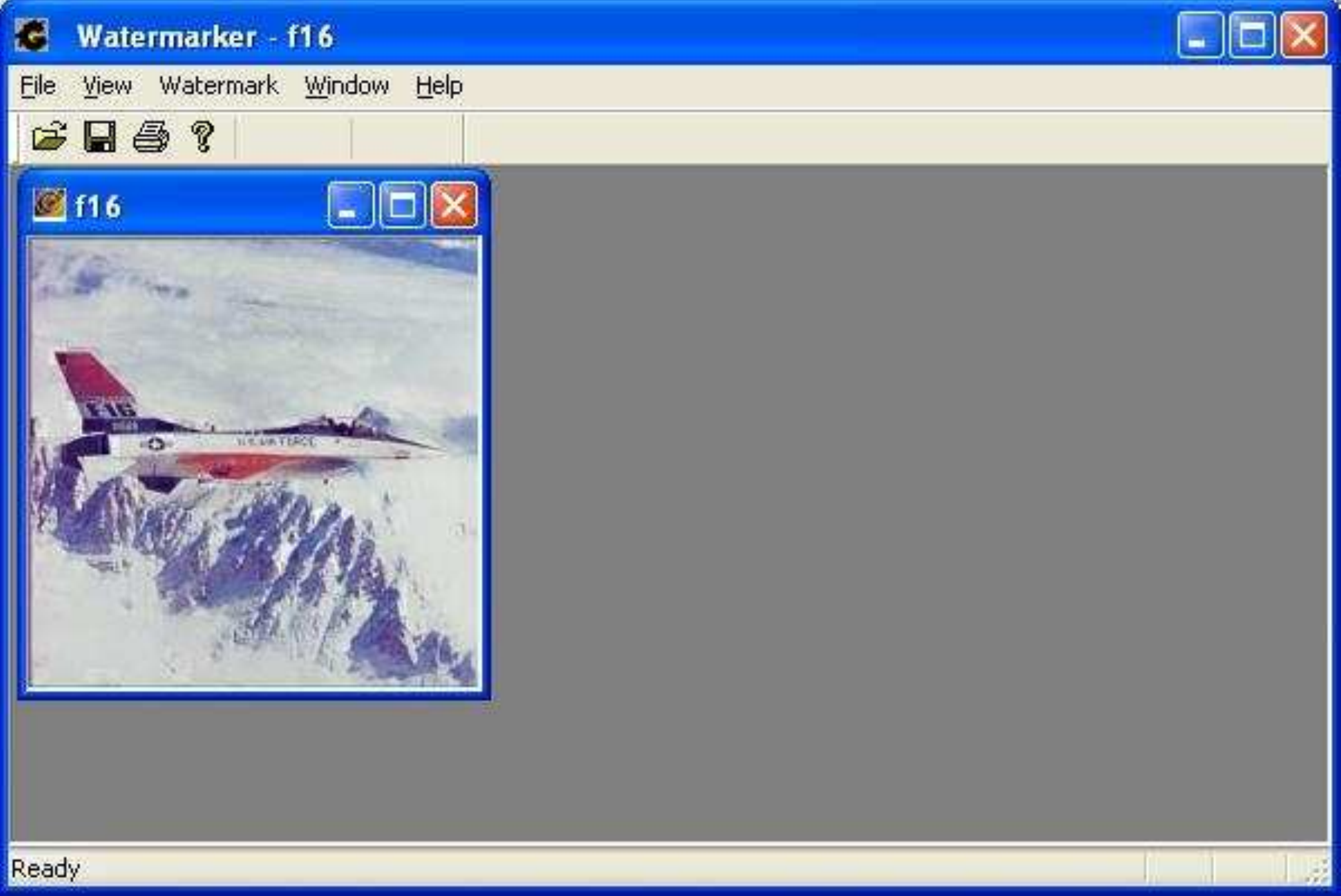} }
\caption{Performing invisible watermark insertion using ISWAR.}
\label{fig:TestingInvisibleInsertionAlgoInISWAR}
\end{figure*}

The invisible watermark embedding operation steps are presented
in Figure~\ref{fig:TestingInvisibleInsertionAlgoInISWAR}.
Here, after opening up, in the same way as above, a host image that is
stored in one of the standard formats, the operation ``Embed a
Non-blind Watermark'' is selected from the ``Watermark'' menu. This
opens up a screen for selecting the watermark image in the
same way as above. This screen has also a provision for typing in a
key of 6-56 characters. This key can contain any text. It is used
for authentication as well as for extraction at the receiver end.
After specifying these two parameters required for invisible
watermark embedding, the ``Embed'' button on the screen may be
clicked. This will trigger execution of the algorithm and the
watermarked image will be opened in the document window.
Simultaneously, the dialog box for embedding invisible watermark
will close.

\begin{figure*}[t]
\centering
\subfigure[Open an image.]
{\label{fig:Image_open_iex}
\includegraphics[height=5.2cm] {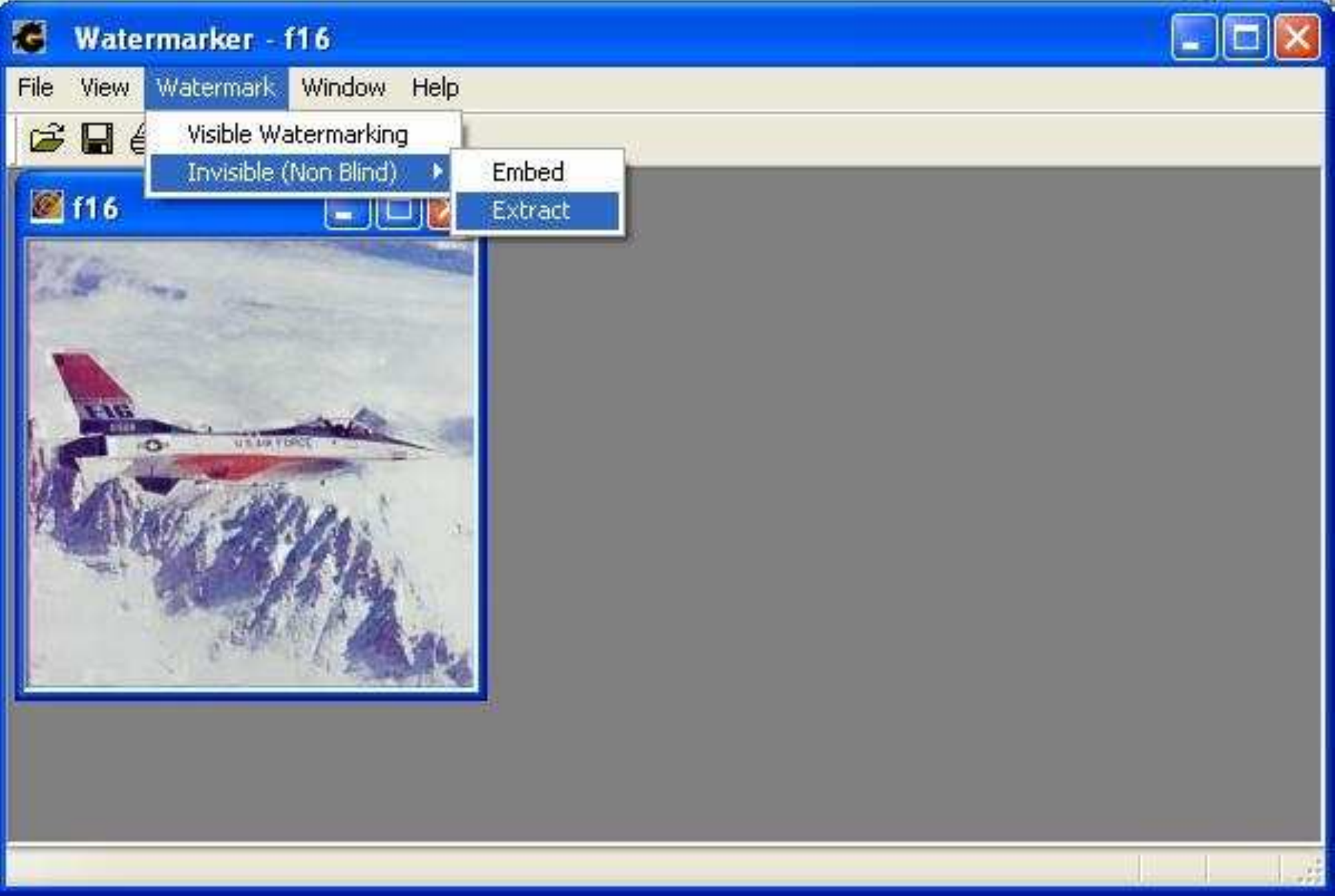} }
\subfigure[Select the watermark. (The same binary image used during
embedding.) Enter the same watermark key which was used during
embedding.]  { \label{fig:invisible_extract_parameters}
\includegraphics[height=5.2cm] {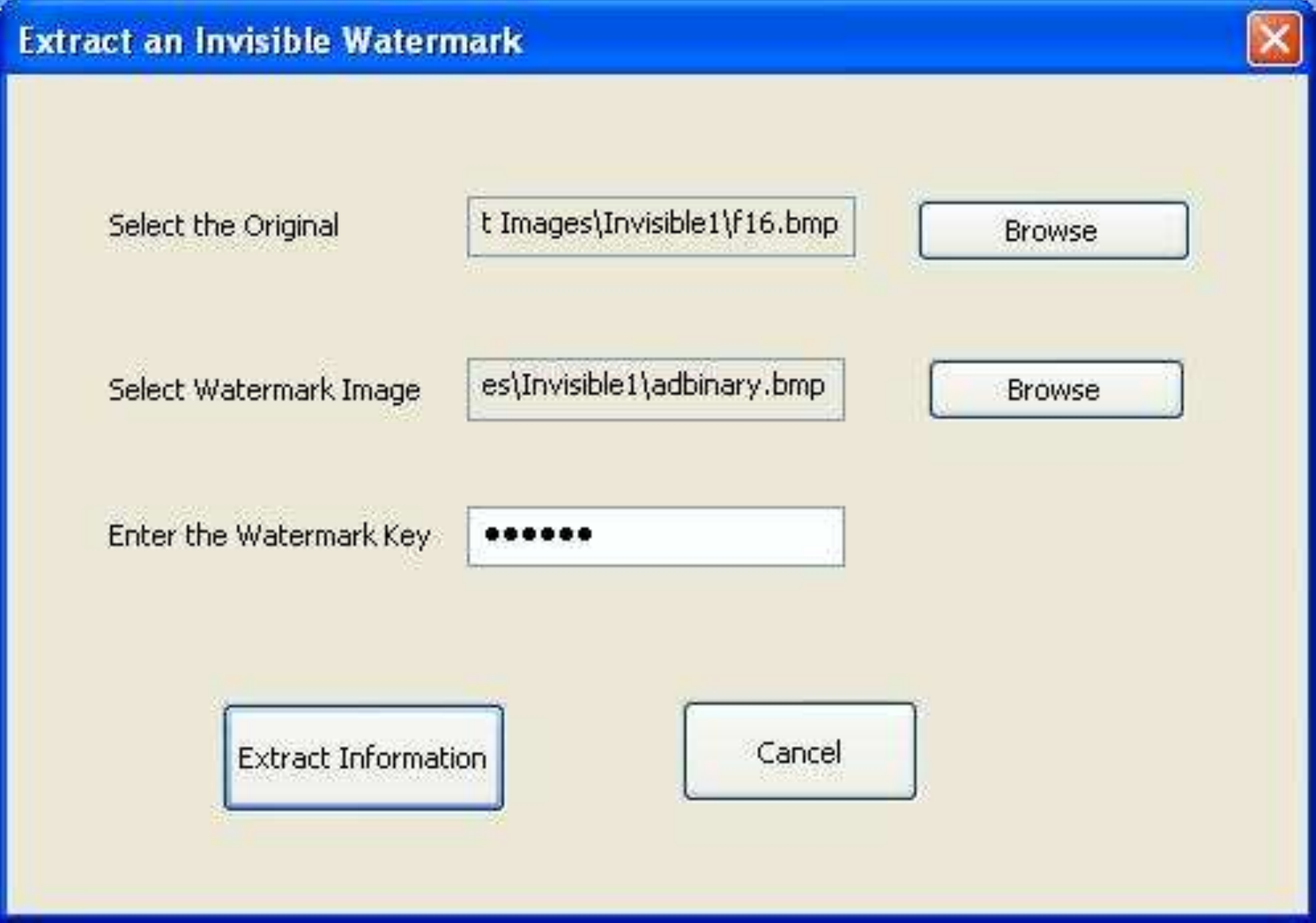} }
\subfigure[Extraction performed: key matched; authentication
successful.] { \label{fig:invisible_extract1}
\includegraphics[height=2.8cm] {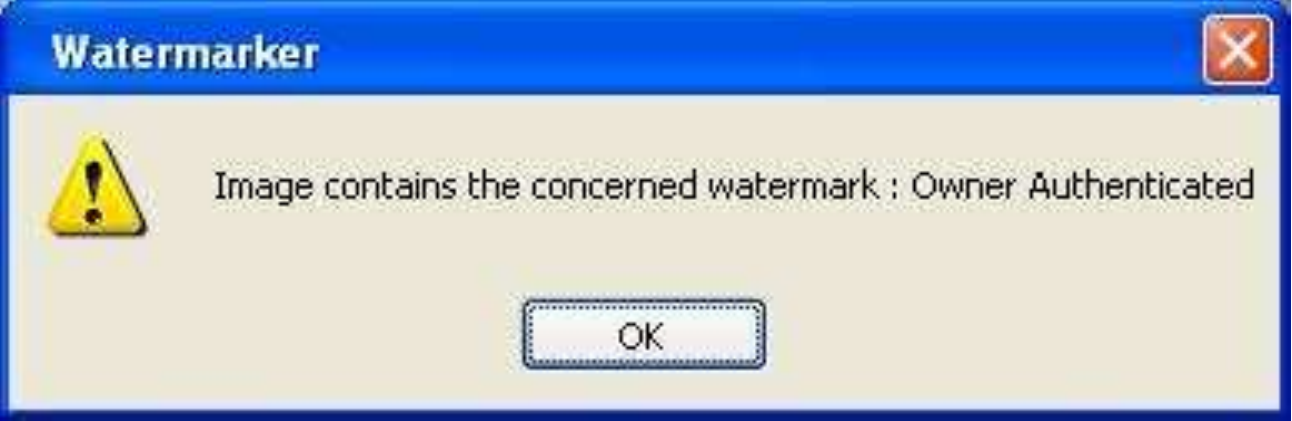} }
\subfigure[Extraction performed: Key did not match; authentication
failure.] { \label{fig:invisible_extract2}
\includegraphics[height=2.8cm] {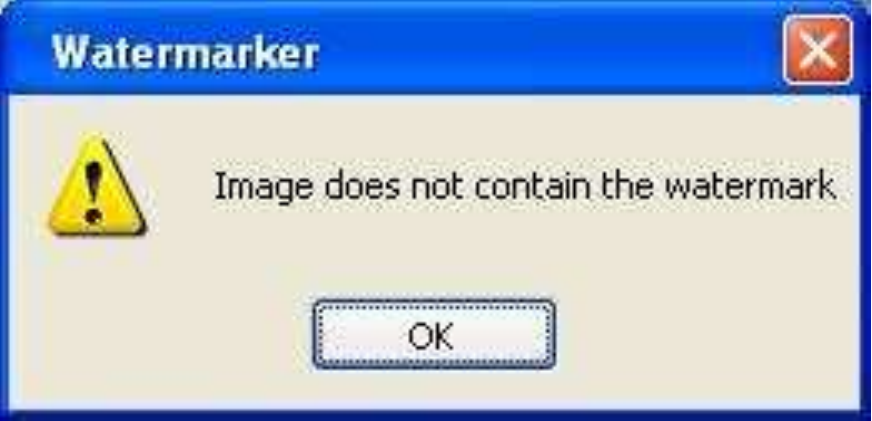} }
\caption{Performing invisible watermark extraction using ISWAR.}
\label{fig:TestingInvisibleExtractionAlgoInIswar}
\end{figure*}

The process for invisible watermark extraction and image
authentication is depicted in Figure~\ref{fig:TestingInvisibleExtractionAlgoInIswar}.
In a watermark extraction operation, the watermarked
image is first opened as in the previous two
operations. Selection of the operation ``Extract an Invisible
Watermark'' from ``Watermark'' menu then opens up a screen for
specification of the required parameters.  On this screen, the user
can specify the  original host image, original binary watermark
image that was embedded,  and the key that was used during the
embedding process. After these parameters are specified correctly,
clicking of the ``Extract'' button on the screen will initiate the
extraction and authentication process that would produce a message
indicating whether the image is authenticated or not.

\section{Performance of ISWAR's Algorithms}
\label{sec:AlgoPerformance}

\begin{figure*}[t]
\centering
\subfigure[Original Lena] {\label{fig:lena_original}
\includegraphics[height=4.2cm] {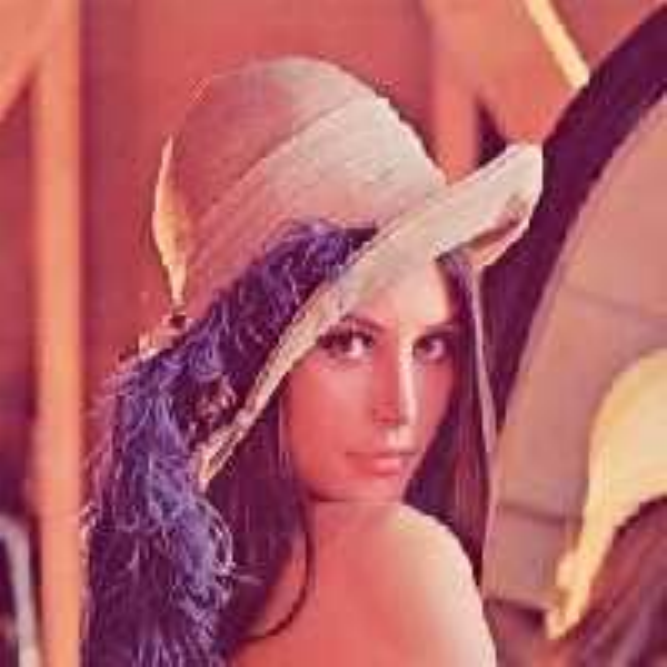} }
\subfigure[With visible watermark (\emph{i} = 10).]  {
\label{fig:lena_visible_unt_logo_strength_10}
\includegraphics[height=4.2cm]
{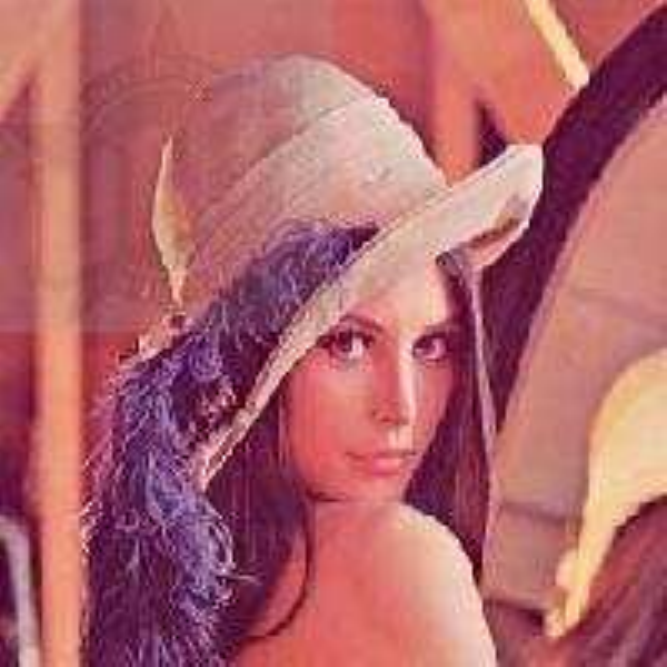} } \subfigure[With invisible
watermark] { \label{fig:lena_invisible}
\includegraphics[height=4.2cm] {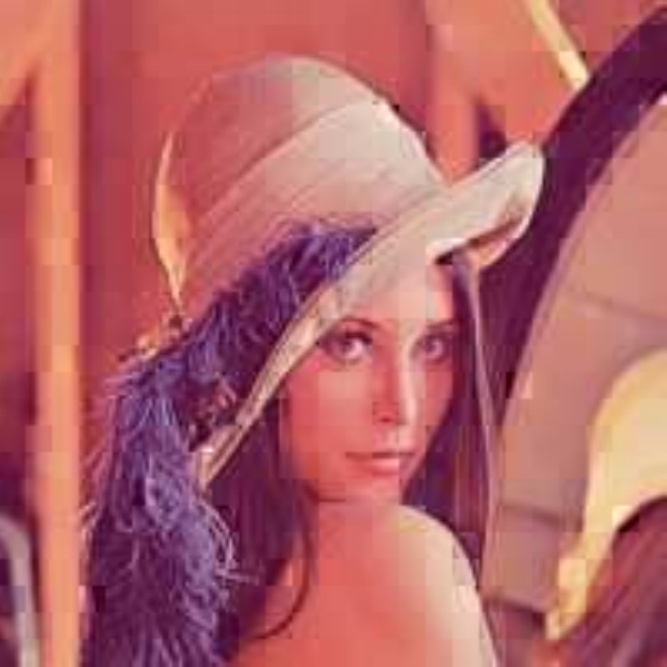} }
\subfigure[Original F16] {\label{fig:f16_original}
\includegraphics[height=4.2cm] {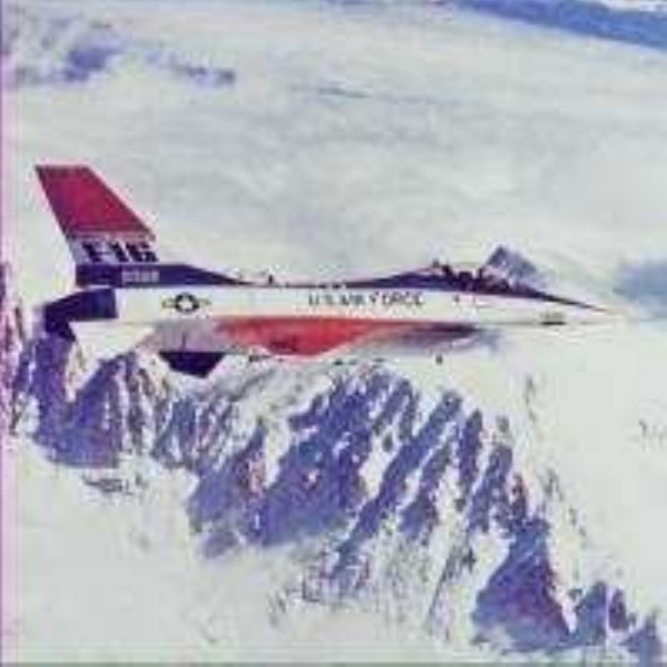} }
\subfigure[With visible watermark (\emph{i} = 10).]  {
\label{fig:f16_visible_unt_logo_strength_10}
\includegraphics[height=4.2cm] {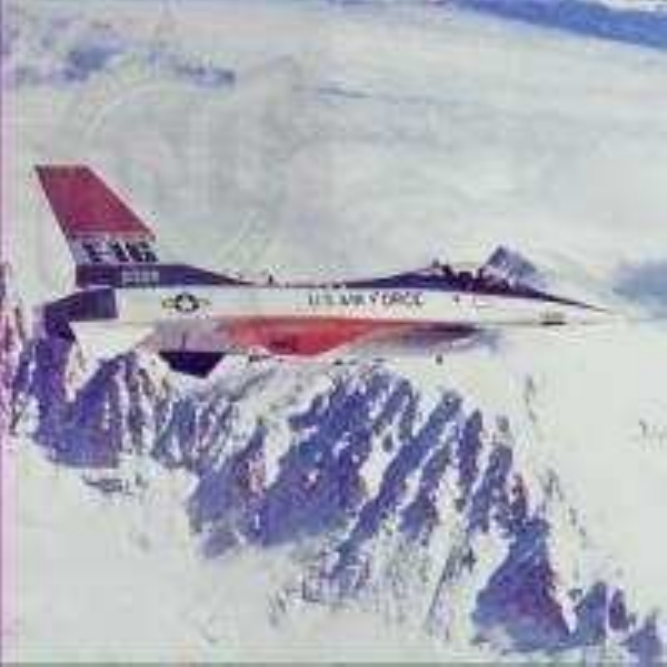}
} \subfigure[With invisible watermark] { 
\includegraphics[height=4.2cm] {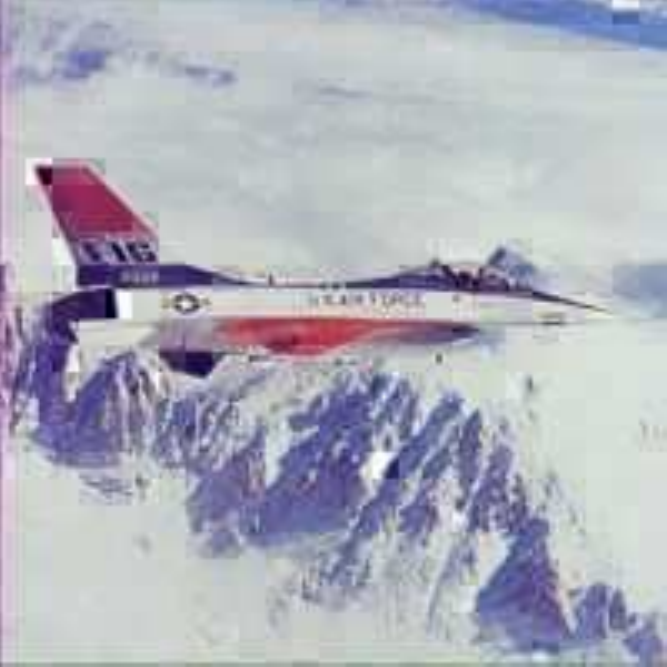} }
\caption{Algorithm evaluation for benchmark image set - 1. The intensity level of the visible watermark is refereed to as $i$.}
\label{fig:VisibleAndInvisibleAlgorithmTesting_1}
\end{figure*}

\begin{figure*}[t]
\centering
\subfigure[Original mandril.] {\label{fig:mandril_original}
\includegraphics[height=4.2cm] {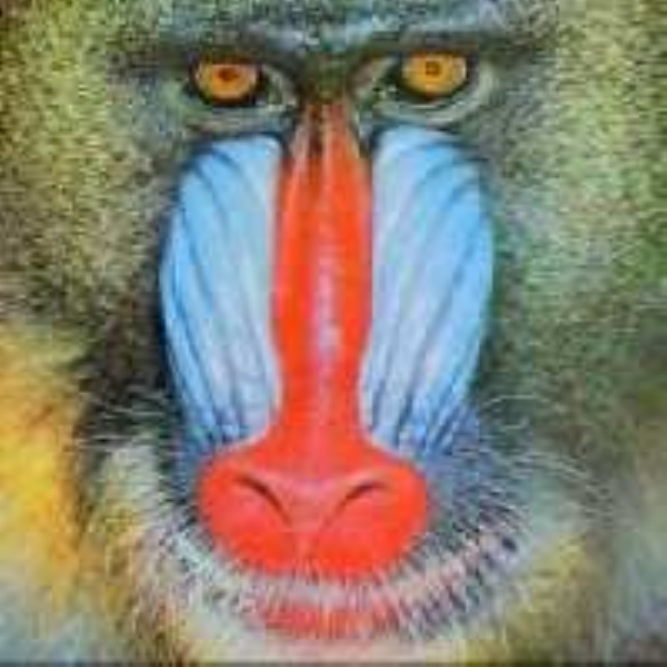} }
\subfigure[With visible watermark (\emph{i}= 20).]  {
\label{fig:mandril_visible_unt_logo_strength_20}
\includegraphics[height=4.2cm]
{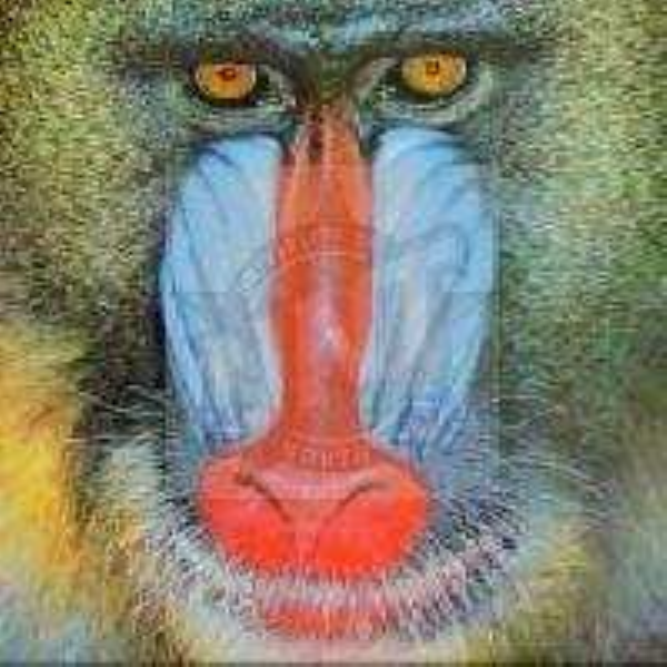} } \subfigure[With
invisible watermark.] { \label{fig:mandril_invisible}
\includegraphics[height=4.2cm] {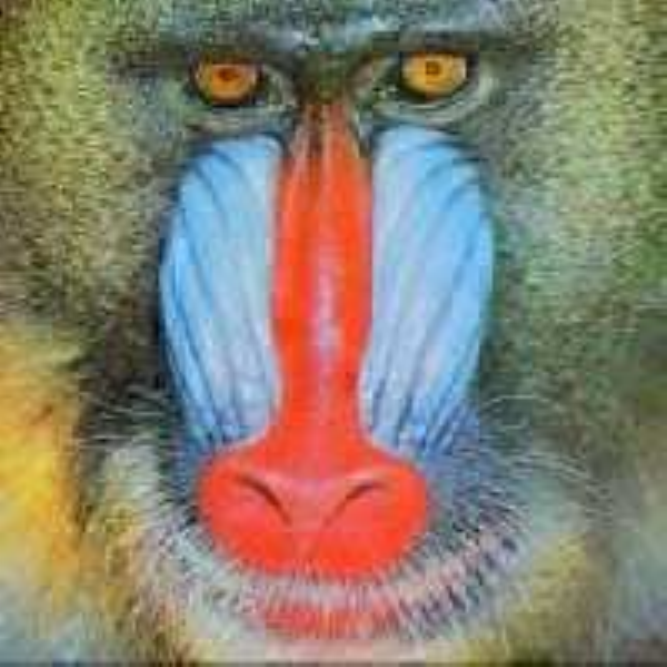} }
\subfigure[Original pepper.] {\label{fig:pepper_original}
\includegraphics[height=4.2cm] {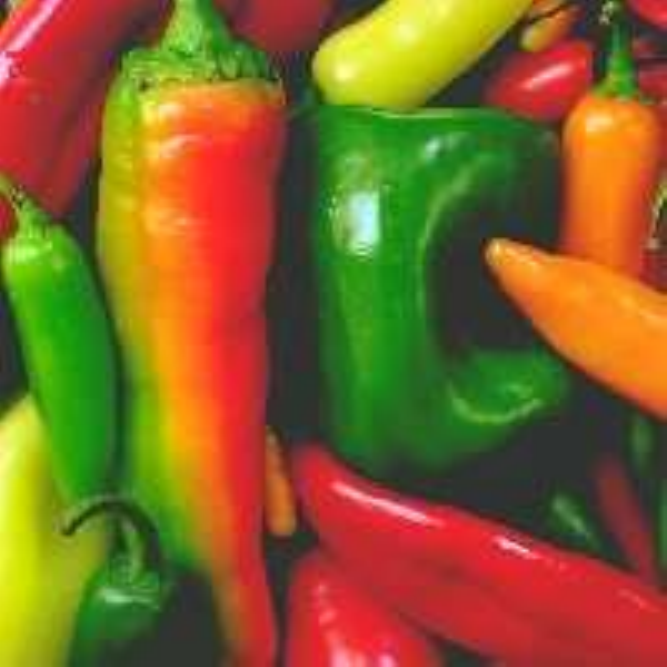} }
\subfigure[With visible watermark (\emph{i} = 3).]  {
\label{fig:pepper_visible_unt_logo_strength_3}
\includegraphics[height=4.2cm]
{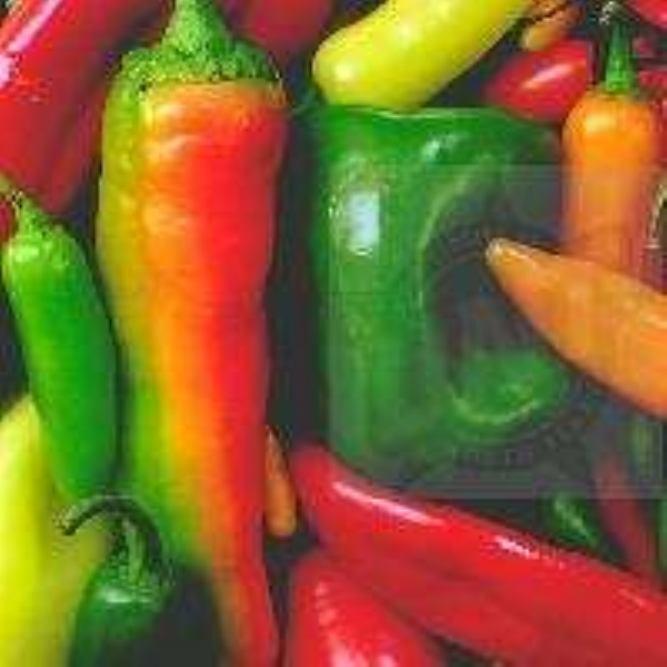} } \subfigure[With invisible
watermark.] { \label{fig:pepper_invisible}
\includegraphics[height=4.2cm] {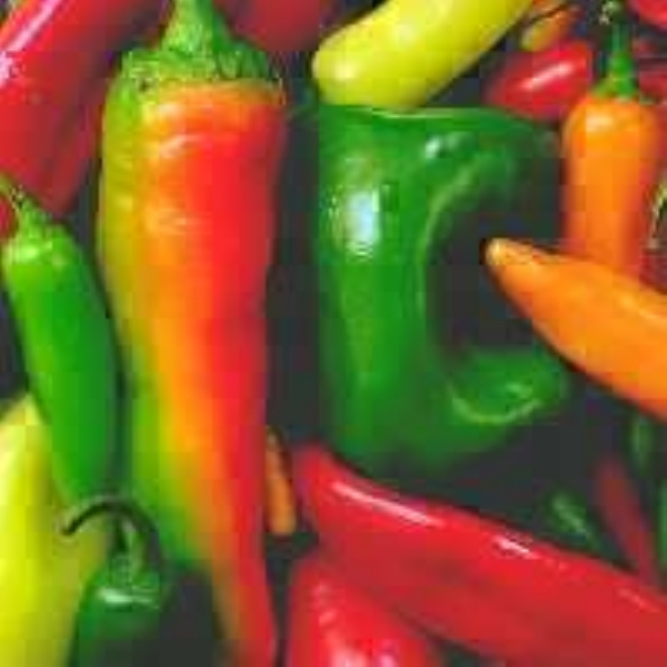} }
\caption{Algorithm evaluation for benchmark image set - 2. The intensity level of the visible watermark is refereed to as $i$.}
\label{fig:VisibleAndInvisibleAlgorithmTesting_2}
\end{figure*}

\begin{figure}[htbp]
\centering
\subfigure[Visible watermark.]
{\label{fig:unt_logo_color_100x100}
\includegraphics[height=3.0cm] {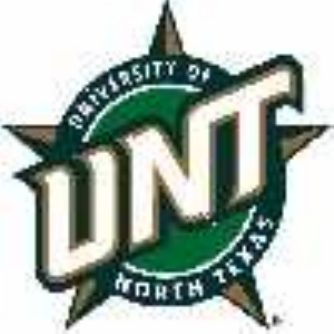} }
\subfigure[Invisible watermark.]  {
\label{fig:invisible_binary_watermark}
\includegraphics[height=1.2cm] {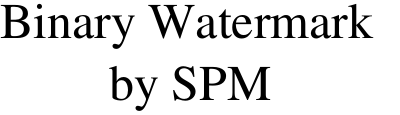} }
\caption{Images used as watermarks.}
\label{fig:WatermarkImagesUsedForTesting}
\end{figure}

The visible and invisible watermarking algorithms are tested
for several benchmark images and different watermarks. For brevity,
the consolidated results of our visible and invisible watermarking
algorithms on the selected benchmark images (Lena, F16, mandrill and pepper) are presented in Figures~\ref{fig:VisibleAndInvisibleAlgorithmTesting_1}
and \ref{fig:VisibleAndInvisibleAlgorithmTesting_2}.
The visible watermark embedding was tested for
various intensity levels of the watermark.
The colored watermark image used for testing of our visible
watermarking algorithm is presented in Figure~\ref{fig:unt_logo_color_100x100}. The binary image used for
invisible watermark embedding and extraction is depicted in Figure~\ref{fig:invisible_binary_watermark}.

The quality of the images obtained using our visible and invisible
watermarking algorithm  may be assessed by visual inspection of the
watermarked images in  Figures~\ref{fig:VisibleAndInvisibleAlgorithmTesting_1}
and \ref{fig:VisibleAndInvisibleAlgorithmTesting_2}. In addition,
the PSNR of the watermarked images provides a quantitative assessment of the quality
of those images. Image quality metrics such as Mean Square Error (MSE) and
PSNR \cite{MohantyJSA2009Oct, MohantyJSS2011} are
applied to quantify the performance of the algorithms.
The MSE and PSNR \cite{MohantyJSA2009Oct, MohantyJSS2011}
are expressed by the following equations:
\begin{eqnarray}
PSNR  & = & {10}\log_{10} \left( \frac{({{2}^{i}-{1}})^{2}}{MSE} \right). \\
MSE  & = & \left( \frac{\sum^{M}_{m=1} \sum^{N}_{n=1} \sum^{3}_{k=1}
I_{E}\left( m, n, k \right)} {3{M \times N}} \right). \\
I_{E} \left( m, n, k \right) & = & \Large|I \left(m, n, k \right)
    - I' \left(m, n, k \right) \Large|^{2},
\end{eqnarray}
where $m$ is the image pixel row from $1$ to $M$, $n$ is the image pixel
column from $1$ to $N$, and $k$ is the index (1 to 3 for RGB color space)
corresponding to the color plane. $p(m,n,k)$ and $q(m,n,k)$ are the
images' pixels after and before processing, respectively, and $i$ is the
bit length of the image pixel, 8 in RGB systems.

In order to obtain a comparative perspective of the visible watermarking algorithm presented here, other selected algorithms from the current literature are cosidered
\cite{HuangIEEEMM2006Feb, ChenCISE2009, YangTCSVT2009May, FarrugiaMELCON2010}.
The PSNR for various benchmark images
is reported in Table~\ref{Table:Visible_Watermarking_Comparison}. The image quality improvement resulting by the proposed algorithm,
which is quantified as an increase in PSNR, is also reported.
As is evident from the Table, the visible watermarking
algorithm generates high-quality images with $PSNR$
improvement in the range of $32.6 - 87.8\%$ than other state-of-the-art algorithms.

\begin{table*}[htbp]
\caption{Comparative Perspective with Existing Visible Watermarking Algorithm.}
\label{Table:Visible_Watermarking_Comparison}
\begin{center}
\begin{tabular}{|l|r|r|}
\hline
Visible	& PSNR  & \% Improvement by\\
Watermarking	& (in dB) & ISWAR Algorithm \\
\hline \hline
{\bf ISWAR} & Lena - 48.2 dB, F16 - 39.5 dB,  & \\
{\bf Algorithm} & mandril - 47.5 dB, pepper - 49.4 dB & NA \\
\hline
Huang, et al. \cite{HuangIEEEMM2006Feb}	& 9.7 - 27.6 dB (Avg. 19.3 dB) & 58.6\% \\
\hline
Chen, et al. \cite{ChenCISE2009}	&  2.4 - 13.2 dB (Avg. 5.6 dB) & 87.8\% \\
\hline
Yang, et al. \cite{YangTCSVT2009May}	& 23.4 - 42.0 dB (Avg. 29.8 dB) & 35.2\% \\
\hline
Farrugia, et al. \cite{FarrugiaMELCON2010}	& 22.6 - 36.8 dB (Avg. 31.4 dB) & 32.6\% \\
\hline
\end{tabular}
\end{center}
\end{table*}

The performance of ISWAR's current invisible watermarking
algorithm with respect to attack resilience has been established by
the results shown in Table~\ref{Table:ComparisonTableHardwareBasedWM} for the well-known
Stirmark attack against the algorithm.
In the table, we reported only the binary outcomes of different
attacks, that is, whether the watermark extracted has survived in
the sense that it is recognizable as a replica of the original
watermark, or not. Unlike some earlier reports, a comparative
analysis on how well the extracted watermark from the attacked image
correlates with its prototype is not presented for the following three reasons:
(1) The paucity of
color image watermarking algorithms makes such a comparison
infeasible. (2) The very purpose of invisible-robust watermarking is
authentication which is a binary decision. Hence, it is more
meaningful to investigate the number of difficult attack scenarios
the algorithm can handle rather than the accuracy for watermark
extraction. As long as the extracted watermark is recognizable, the
purpose is served. (3) There is always a tradeoff between the
perceptual quality of the watermarked image produced by an algorithm
and the quality of the extracted watermark under noise and other
degradations. Hence, after establishing with different images and
different watermark intensities that the visual quality of our
watermarked images are acceptable, the results that are presented
help benchmarking our algorithm against the ideal algorithm that
survives all the attack types in the Stirmark attack.

\begin{table*}[t]
\caption{Attacks performed using benchmarks for testing of the
invisible-robust algorithm.}
\begin{center}
\begin{tabular}{|l|l|l|l|l|l|}
\hline

 & \multicolumn{4}{|c|}{For Various Benchmark Images} \\

\cline{2-5}

Attacks Performed for Testing & Lena  & F16  & mandril & pepper \\

\hline \hline

JPEG Compression  & Survived  & Survived &
Survived  & Survived \\
\hline

Gray scaling 16 levels  &  Survived  & Survived &
Survived & Survived \\
\hline

Gray scaling 256 levels, JPEG   & Survived  &
Survived  &  Survived  & Survived \\

\hline

Blurring, JPEG Compression  & Survived  & Survived
& Survived &  Survived \\

\hline

Partial cropping  & Survived  & Survived & Survived  &
Survived \\

\hline

Stirmark – Self Similarities  & Survived  & Survived  &
Survived  & Survived \\

\hline

Stirmark – JPEG compression  & Survived  & Survived
 & Survived  & Survived  \\

\hline

Stirmark – median filtering  & Survived  & Survived  & Survived &
Survived   \\

\hline

Stirmark – Random Distortions  & Survived  & Survived   & Survived &
Survived   \\
\hline
\end{tabular}
\end{center}
\label{Table:ComparisonTableHardwareBasedWM}
\end{table*}

In order to obrain a comparative perspective of the performance of the invisible-robust watermarking algorithm, selected watermarking techniques are discussed.
The proposed invisible watermarking algorithm is unique in using
both AC and DC DCT coefficients for watermarking candidates, encryption in the framework, and hides binary images inside the host image by selective addition and subtraction.
Thus, it can allow secure storage of sensitive information including
binary fingerprint and hand signature, in a cover image.
The results are reported in Table~\ref{Table:Invisible_Watermarking_Comparison}.
As it is evident from the Table, the invisible-robust watermarking
algorithm generates high-quality
images with $PSNR$ improvement in the range of $56.7 - 76.7\%$ compared to the state-of-the-art.

\begin{table*}[t]
\caption{Comparative perspective with existing invisible-robust watermarking algorithm.}
\label{Table:Invisible_Watermarking_Comparison}
\begin{center}
\begin{tabular}{|l|r|r|}
\hline
Invisible	& PSNR  & \% Improvement by  \\
Watermarking	& (in dB) & ISWAR Algorithm\\
\hline \hline
{\bf ISWAR} & Lena - 105 dB, F16 - 99 dB,  & \\
{\bf Algorithm} & mandril -101 dB, pepper - 108 dB& NA \\
\hline
Wu, et al. \cite{WuCGI2001}	& 24.9 - 47.9 dB (Avg. 38.3 dB) & 62.4 \%\\
\hline
Pai, et al. \cite{PaiIEICETIS2006Apr}	& 29.1 - 39.6 dB (Avg. 34.7 dB) & 65.9 \%\\
\hline
Saxena, et al. \cite{SaxenaICSPCA2007}	& 23.67 - 23.73 dB (Avg. 23.7 dB) & 76.7 \% \\
\hline
Rao, et al. \cite{RaoIJCSNS2009Mar}	& 42.9 - 45.8 dB (Avg. of 44.1 dB) & 56.7 \%\\
\hline
\end{tabular}
\end{center}
\end{table*}

\section{Summary and Conclusion}
\label{sec:SummaryAndConclusion}

In this paper, an object-oriented secure imaging system
with watermarking called ISWAR is presented along with novel and effective
algorithms for visible and invisible watermarking, extraction and authentication.
The system is useful for diverse image security applications.
The OO design makes it
extensible. Like any active system, ISWAR is evolving and hence
many enhancements are underway. For example, an extension that is
under consideration for the invisible-robust watermarking is to have two
watermarks, a user specific binary watermark and a synthetic
watermark generated by the system, and fuse them together into the
cover for additional protection and better image quality. Other
possible extensions include use of wavelet transforms for embedding
of strong watermarks. Here, Daubechies 9/7 filters will be investigated
compared to Haar
wavelets as the latter may leave out some detail. Blind extraction
of invisible watermarks is also a planned extension particularly
due to its usefulness in authentication at the receiver end as well
as identification of secretive communication among enemies using
steganography.

\section*{Acknowledgements}

The author would like to acknowledge the help of
Elias Kougianos and Parthasarathy Guturu from the
University of North Texas for their inputs
in improving the presentation.
The author would like to acknowledge the help of
Rajan Sheth, Adrain Pinto, and Marina Chandy from
the St. Francis Institute of Technology,
Mumbai - 400103, India, for their help
in some parts of the software implementation.
The preliminary idea of the imaging watermarking system
discussed in this archival journal paper was introduced
in our ISCE 2007 conference presentation \cite{MohantyISCE2007}.



\begin{thebibliography}{10}

\bibitem{BenderIBMSystemsJournal2000}
W.~Bender, W.~Butera, D.~Gruhl, R.~Hwang, F.~J. Paiz, and S.~Pogreb.
\newblock {Applications for Data Hiding}.
\newblock {\em IBM Systems Journal}, 39(3 and 4):547--568, 2000.

\bibitem{BenderIBMSystemsJournal1996}
W.~Bender, D.~Gruhl, and N.~Morimoto.
\newblock {Techniques for Data Hiding}.
\newblock {\em IBM Systems Journal}, 35(3):313--336, 1996.

\bibitem{ChenICCE2000}
Ding-Yun Chen and Chun-Hsiang Huang Ja-Ling Wu~Ming Ouhyoung.
\newblock A shift-resisting blind watermark system for panoramic images.
\newblock In {\em Proceedings of International Conference on Consumer
  Electronics}, pages 8--9, 2000.

\bibitem{ChenCISE2009}
J.~J. Chen, T.~M. Ng, A.~Lakshminarayanan, and H.~K. Garg.
\newblock Adaptive visible watermarking using otsu's thresholding.
\newblock In {\em Proceedings of International Conference on Computational
  Intelligence and Software Engineering}, pages 1--4, 2009.

\bibitem{CoxJASP2002}
I.~J. Cox and M.~L. Miller.
\newblock {Electronic Watermarking : The First 50 Years}.
\newblock {\em EURASIP Journal of Applied Signal Processing}, 2002(2):126--132,
  February 2002.

\bibitem{EmmanuelMMSJ2003}
S.~Emmanuel and M.~S. Kankanhalli.
\newblock {A Digital Rights Management Scheme for Broadcast Video}.
\newblock {\em ACM-Springer Verlag Multimedia Systems Journal}, 8(6):444--458,
  June 2003.

\bibitem{EskiciogluSP2001}
A.~M. Eskicioglu and E.~J. Delp.
\newblock {An Overview of Multimedia Content Protection in Consumer Electronics
  Devices}.
\newblock {\em Elsevier Signal Processing : Image Communication}, 16:681--699,
  2001.

\bibitem{FarrugiaMELCON2010}
R.~A. Farrugia.
\newblock A reversible visible watermarking scheme for compressed images.
\newblock In {\em Proceedings of 15th IEEE Mediterranean Electrotechnical
  Conference}, pages 212--217, 2010.

\bibitem{FrattolilloISIAS2008}
F.~Frattolillo and F.~Landolfi.
\newblock Designing a drm system.
\newblock In {\em Proceedings of Fourth International Conference on Information
  Assurance and Security}, pages 221--226, 2007.

\bibitem{GranrathProceedingsIEEE1981}
D.~J. Granrath.
\newblock {The Role of Human Visual Models in Image Processing}.
\newblock {\em Proceedings of the IEEE}, 69(5):552--561, May 1981.

\bibitem{GuoMMSJ2003}
H.~Guo and N.~D. Georganas.
\newblock {A Novel Approach to Digital Image Watermarking Based on a
  Generalized Secret Sharing Scheme}.
\newblock {\em ACM-Springer Verlag Multimedia Systems Journal}, 9(3):228--238,
  March 2003.

\bibitem{HuKwongISCS2004}
Y.~Hu and S.~Kwong.
\newblock {An Image Fusion Based Visible Watermarking Algorithm}.
\newblock In {\em Proceedings of the 2003 International Symposium on Circuits
  and Systems, 2003. ISCAS '03}, volume~3, pages III--794 -- III--797, 25-28
  May 2003.

\bibitem{HuaICPRIP2001}
X.~S. Hua, J.~F. Feng, and Q.~Y. Shi.
\newblock {Public Multiple Watermarking Resistant to Cropping}.
\newblock In {\em Proceedings of the 6th international conference on pattern
  recognition and information processing}, pages 263--268, 2001.

\bibitem{HuangIEEEMM2006Feb}
Biao-Bing Huang and Shao-Xian Tang.
\newblock {A contrast-sensitive visible watermarking scheme}.
\newblock {\em IEEE Multimedia}, 13(2):60--66, February 2006.

\bibitem{KankanhalliICMCS1999}
M.~S. Kankanhalli, Rajmohan, and K.~R. Ramakrishnan.
\newblock {Adaptive Visible Watermarking of Images}.
\newblock In {\em Proceedings of the IEEE International Conference on
  Multimedia Computing Systems}, pages 9568--9573, 1999.

\bibitem{KougianosIJCEE2009Mar}
Elias Kougianos, Saraju~P. Mohanty, and Rabi~N. Mahapatra.
\newblock Hardware assisted watermarking for multimedia.
\newblock {\em Elsevier International Journal on Computers {\&} Electrical
  Engineering}, 35(2):339--358, 2009.

\bibitem{KundurIEEEProceedings2004}
D.~Kundur and K.~Karthik.
\newblock {Digital Fingerprinting and Encryption Principles for Digital Rights
  Management}.
\newblock {\em Proceedings of the IEEE Special Issue on Enabling Security
  Technologies for Digital Rights Management}, 92:918--932, 2004.

\bibitem{LiuCSSE2008}
Quan Liu and Hong Liu.
\newblock An intelligent digital right management system based on multi-agent.
\newblock In {\em Proceedings of International Conference on Computer Science
  and Software Engineering}, pages 505--507, 2008.

\bibitem{LuminiICME2004}
A.~Lumini and D.~Maio.
\newblock Adaptive positioning of a visible watermark in a digital image.
\newblock In {\em Proceedings of IEEE International Conference on Multimedia
  and Expo}, pages 967--970, 2004.

\bibitem{MasonIBC2000}
A.~J. Mason, R.~A. Salmon, O.~H. Werner, and J.~E. Devlin.
\newblock {User Requirements for Watermarking in Broadcast Applications}.
\newblock In {\em Proceedings of the IEE International Broadcasting Convention
  (IBC)}, 2000.

\bibitem{MintzerIBMJournal1996}
F.~C. Mintzer, L.~E. Boyle, A.~N. Cazes, B.~S. Christian, S.~C. Cox, F.~P.
  Giordano, H.~M. Gladney, J.~C. Lee, M.~L. Kelmanson, A.~C. Lirani, K.~A.
  Magerlein, A.~M.~B. Pavani, and F.~Schiattarella.
\newblock {Towards online Worldwide Access to Vatican Library Materials}.
\newblock {\em IBM Journal of Research and Development}, 40(2):139--162, Mar
  1996.

\bibitem{MohantyJSA2009Oct}
S.~P. Mohanty.
\newblock A secure digital camera architecture for integrated real-time digital
  rights management.
\newblock {\em Elsevier Journal of Systems Architecture - Embedded Systems
  Design}, 55(10-12):468--480, 2009.

\bibitem{MohantyJSS2011}
S.~P. Mohanty and E.~Kougianos.
\newblock Real-time perceptual watermarking architectures for video
  broadcasting.
\newblock {\em Elsevier Journal of Systems and Software}, 84(5):724--738, May
  2011.

\bibitem{MohantyACMMM1999}
S.~P. Mohanty, K.~R. Ramakrishnan, and M.~S. Kankanhalli.
\newblock {A Dual Watermarking Technique for Images}.
\newblock In {\em Proceedings of the 7th ACM International Multimedia
  Conference (Vol. 2)}, pages 49--51, 1999.

\bibitem{MohantyICME2000}
S.~P. Mohanty, K.~R. Ramakrishnan, and M.~S. Kankanhalli.
\newblock {A DCT Domain Visible Watermarking Technique for Images}.
\newblock In {\em Proceedings of the IEEE International Conference on
  Multimedia and Expo}, pages 1029--1032, 2000.

\bibitem{MohantyISCE2007}
S.~P. Mohanty, R.~Sheth, A.~Pinto, and M.~Chandy.
\newblock {CryptMark: A Novel Secure Invisible Watermarking Technique for Color
  Images}.
\newblock In {\em Proceedings of the 11th IEEE International Symposium on
  Consumer Electronics}, pages 1--6, 2007.

\bibitem{Mohanty_JSA2009Oct}
S.P. Mohanty.
\newblock {A Secure Digital Camera Architecture for Integrated Real-Time
  Digital Rights Management}.
\newblock {\em Journal of Systems Architecture}, 55(10-12):468--480, 2009.

\bibitem{Mohanty_CDT2007Sep}
S.P. Mohanty, E.~Kougianos, and N.~Ranganathan.
\newblock {VLSI Architecture and Chip For Combined Invisible Robust and Fragile
  Watermarking}.
\newblock {\em Computers \& Digital Techniques, IET}, 1(5):600--611, 2007.

\bibitem{Mohanty_ICMPS2000}
S.P. Mohanty, KR~Ramakrishnan, and M.S. Kankanhalli.
\newblock {An Adaptive DCT Domain Visible Watermarking Technique for Protection
  of Publicly Available Images}.
\newblock In {\em Proceedings of the International Conference on Multimedia
  Processing and Systems}, pages 195--198. Citeseer, 2000.

\bibitem{NikolaidisICIP2001}
A.~Nikolaidis, S.~Tsekeridou, A.~Tefas, and V.~Solachidis.
\newblock {A Survey on Watermarking Application Scenarios and Related Attacks}.
\newblock In {\em Proceedings of the IEEE International Conference on Image
  Processing}, volume~1, pages 991--994, 2001.

\bibitem{PaiIEICETIS2006Apr}
Y.~T. Pai and S.~J. Ruan.
\newblock {Low Power Block-Based Watermarking Algorithm}.
\newblock {\em IEICE Transactions on Information and Systems},
  E89-D(4):1507--1514, 2006.

\bibitem{RaoIJCSNS2009Mar}
N.~N. Rao, P.~Thrimurthy, and B.~R. Babu.
\newblock {A Novel Scheme for Digital Rights Management of Images Using
  Biometrics}.
\newblock {\em International Journal of Computer Science and Network Security},
  9(3):157--167, March 2009.

\bibitem{SaxenaICSPCA2007}
V.~Saxena and J.~P. Gupta.
\newblock Collusion attack resistant watermarking scheme for images using dct.
\newblock In {\em Proceedings of 15th IEEE Conf. on Signal Processing and
  Communications Applications}, pages 1--4, 2007.

\bibitem{SheppardACMMM2001}
N.~P. Sheppard, R.~S. Naini, and P.~Ogunbona.
\newblock {On Multiple Watermarking}.
\newblock In {\em Proceedings of the ACM Multimedia workshops on multimedia and
  security: new challenges}, pages 3--6, 2001.

\bibitem{SungSERA2006}
Kyung-Sang Sung, Seung-Heon Lee, Bo-Hyun Wang, and Hae-Seok Oh.
\newblock Digital watermark system based on improved security through
  pre-processing of watermarked data using information of image discrete
  frequency.
\newblock In {\em Proceedings of Fourth International Conference on Software
  Engineering Research}, pages 276--280, 2006.

\bibitem{TopkaraEtAlLecNotesCompSc2005}
M.~Topkara, A.~Kamara, M.~Atallah, and C.~Nita-Rotaru.
\newblock {ViWiD: Visible Watermark Based Defense Against Phishing}.
\newblock {\em Lecture Notes in Computer Science (LNCS), IWDW 2005},
  3710:470--484, 2005.

\bibitem{TsaiICME2007}
Han-Min Tsai and Long-Wen Chang.
\newblock {A High Secure Reversible Visible Watermarking Scheme}.
\newblock In {\em IEEE International Conference on Multimedia and Expo}, pages
  2106--2109, 2007.

\bibitem{VoloshynovskiyIEEECommMag2001}
S.~Voloshynovskiy, S.~Pereira, T.~Pun, J.J. Eggers, and J.K. Su.
\newblock {Attacks on Digital Watermarks: Classification, Estimation-based
  Attacks and Benchmarks}.
\newblock {\em IEEE Communications Magazine}, 39(9):118--126, August 2001.

\bibitem{VoyatzisgProceedingsIEEE1999}
G.~Voyatzis and I.~Pitas.
\newblock {The Use of Watermarks in the Protection of Digital Multimedia
  Products}.
\newblock {\em Proceedings of the IEEE}, 87(7):1197--1207, July 1999.

\bibitem{WuCGI2001}
Y.~Wu, X.~Guan, and M.~S. Kankanhalli.
\newblock Robust invisible watermarking of volume data using the 3d dct.
\newblock In {\em Proc. of Computer Graphics International (CGI)}, pages
  359–--362, 2001.

\bibitem{XieSNPD2007}
Rongsheng Xie, Keshou Wu, Jiangbo Du, and Chunguang Li.
\newblock Survey of public key digital watermarking systems.
\newblock In {\em Proceedings of Software Engineering, Artificial Intelligence,
  Networking, and Parallel/Distributed Computing}, pages 439--443, 2007.

\bibitem{YangTCSVT2009May}
Ying Yang, Xingming Sun, Hengfu Yang, Chang-Tsun Li, and Rong Xiao.
\newblock {A Contrast-Sensitive Reversible Visible Image Watermarking
  Technique}.
\newblock {\em IEEE Transactions on Circuits and Systems for Video Technology},
  19(5):656--667, 2009.

\end{thebibliography}
\end{document}